%% file: gama_compacts.tex
\documentclass{aa}  

\usepackage{graphicx}
\usepackage{txfonts}
\usepackage{color}

\begin{document}

   \title{Galaxy and Mass Assembly (GAMA): Accurate number densities \& environments of massive ultracompact galaxies at 0.02 < z < 0.3}

   \subtitle{}

   \author{F. Buitrago\inst{1,2} \fnmsep \thanks{Based on observations made with ESO Telescopes at the La Silla Paranal Observatory under programme ID 179.A-2004.} \fnmsep \thanks{Based on observations made with ESO Telescopes at the La Silla Paranal Observatory under programme ID 177.A-3016.}
   \and I. Ferreras\inst{3}
   \and L.~S. Kelvin\inst{4}
   \and I.~K. Baldry\inst{4}
   \and L. Davies\inst{5}
   \and J. Angthopo\inst{3}
   \and S. Khochfar\inst{6}
   \and A.~M. Hopkins\inst{7}
   \and S.~P. Driver\inst{5}
   \and S. Brough\inst{8}
   \and J. Sabater\inst{6}
   \and C.~J. Conselice\inst{9}
   \and J. Liske\inst{10}
   \and B.~W. Holwerda\inst{11}
   \and M.~N. Bremer\inst{12}
   \and S. Phillipps\inst{12}
   \and \'A.~R. L\'opez-S\'anchez\inst{7,13} 
   \and A.~W. Graham\inst{14}}
\institute{Instituto de Astrof\'{\i}sica e Ci\^{e}ncias do Espa\c{c}o, Universidade de Lisboa, OAL, Tapada da Ajuda, PT1349-018 Lisbon, Portugal.\\\email{fbuitrago@oal.ul.pt}
\and
Departamento de F\'{i}sica, Faculdade de Ci\^{e}ncias, Universidade de Lisboa, Edif\'{i}cio C8, Campo Grande, PT1749-016 Lisbon, Portugal.
\and
Mullard Space Science Laboratory, University College London, Holmbury St Mary, Dorking, Surrey RH5 6NT, UK.\\
\email{i.ferreras@ucl.ac.uk}
\and
Astrophysics Research Institute, Liverpool John Moores University, IC2, Liverpool Science Park, 146 Brownlow Hill, Liverpool L3 5RF, UK.
\and
ICRAR, School of Physics, University of Western Australia, 35 Stirling Highway, Crawley, WA 6009, Australia.
\and
Institute for Astronomy, University of Edinburgh, Royal Observatory, Edinburgh EH9 3HJ, UK.
\and
Australian Astronomical Observatory, 105 Delhi Rd, North Ryde, NSW 2113, Australia.
\and
School of Physics, University of New South Wales, NSW 2052, Australia.
\and 
Centre for Astronomy and Particle Theory, School of Physics \& Astronomy, University of Nottingham, Nottingham, NG7 2RD, UK
\and
Hamburger Sternwarte, Universit\"at Hamburg, Gojenbergsweg 112, D-21029 Hamburg, Germany.
\and
Department of Physics and Astronomy, 102 Natural Science Building, University of Louisville, Louisville KY 40292, USA.
\and 
H.~H. Wills Physics Laboratory, University of Bristol, Tyndall Avenue, Bristol, BS8 1TL, UK.
\and 
Department of Physics and Astronomy, Macquarie University, NSW 2109, Australia
\and
Centre for Astrophysics and Supercomputing, Swinburne University of Technology, Victoria 3122, Australia
}

   \date{}

 
  \abstract
   {Massive Ultracompact Galaxies (MUGs) are common at z$=$2-3, but very rare in the nearby Universe. Simulations predict that the few surviving MUGs should reside in galaxy clusters, whose large relative velocities prevent them from merging, thus maintaining their original properties (namely stellar populations, masses, sizes and dynamical state).}
   {Our goal is to obtain a complete census of the MUG population at 0.02$<$z$<$0.3, determining the number density, population properties and environment.}
   {We take advantage of the high-completeness, large-area spectroscopic GAMA survey, complementing it with deeper imaging from the KiDS and VIKING surveys. We find a set of 22 bona-fide MUGs, defined as having high stellar mass ($>$8$\times$10$^{10}$M$_\odot$) and compact size (R$_{\rm e}<$2\,kpc). An additional set of 7 lower-mass objects (6$\times$10$^{10}$ $<$ M$_{\star}$/M$_{\odot}$ $<$ 8$\times$10$^{10}$) are also potential candidates according to typical mass uncertainties.}
   {The comoving number density of MUGs at low redshift (z $<$ 0.3) is constrained at $(1.0\pm 0.4)\times 10^{-6}$\,Mpc$^{-3}$, consistent with galaxy evolution models. However, we find a mixed distribution of old and young galaxies, with a quarter of the sample representing (old) relics. MUGs have a predominantly early/swollen disk morphology (S\'ersic index 1$<n<$2.5) with high stellar surface densities ($\langle\Sigma_{\rm e}\rangle$ $\sim$ 10$^{10}$ M$_{\odot}$ Kpc$^{-2}$). Interestingly, a large fraction feature close companions -- at least in projection --  suggesting that many (but not all) reside in the central regions of groups. Halo masses show these galaxies inhabit average-mass groups. }
{As MUGs are found to be almost equally distributed among environments of different masses, their relative fraction is higher in more massive overdensities, matching the expectations that some of these galaxies fell in these regions at early times. However, there must be another channel leading some of these galaxies to an abnormally low merger history because our sample shows a number of objects that do not inhabit particularly dense environments.}


   \titlerunning{Number densities and environments for low-z Massive Ultracompact Galaxies}
   \authorrunning{Buitrago et al.}

   \maketitle
%

\section{Introduction}

Massive galaxies -- defined as having stellar masses exceeding the characteristic mass of the galaxy mass function (M$^{\star}$), or roughly those with $\gtrsim$10$^{11}$M$_{\odot}$ \citep[see e.g.][]{Baldry12,Kelvin14}-- are privileged testbeds for galaxy evolution theories, as they generally underwent dramatic changes in their sizes, morphologies and star formation rates at an accelerated pace in comparison with lower mass objects across cosmic time (e.g. \citealt{Daddi05,Trujillo06a,Trujillo07,Buitrago08,vanDokkum10,I3,Conselice11,Bruce12,Ferreras12,Buitrago13,McLure13a,Driver:13,VanderWel14,Buitrago14} to name but a few; see also \citealt{Khochfar06,Naab09,Oser10,Johansson12,Nipoti12,Zolotov15,Wellons16,Genel18,Lapi18} from the theoretical point of view). Historically, a number of nearby massive galaxies were claimed to be rather compact as far back as \citet{ZwKow:68}, tentatively ascribing such a trend to cluster-related mechanisms \citet{Strom:78}.


At $z =$ 2-3, the majority of massive galaxies display very small sizes (effective radii $r_{\rm e}$ $\leq$ 1-2 kpc) and thus they were branded ``red nuggets" \citep{Damjanov09}, with the term ``red" adopted as some of them are already passive even at such early stages of cosmic evolution. It was also claimed that the nearby Universe was devoid of such ultracompact massive galaxies \citep[e.g.][]{Cimatti08}. \citet{Trujillo09}, later confirmed by \citet{Taylor10} and \citet{Shih11}, found a tiny (in terms of number density) population of such galaxies at $z <$ 0.2. Nevertheless, the stellar populations of these newly found galaxies were rather young ($\sim$2 Gyr) and metal rich (Z$\sim$Z$_{\odot}$), meaning that these objects were not survivors from the $z >$ 2 compact massive galaxy population \citep{Ferre-Mateu13}. However, these results raised hopes of finding the local analogs of high-z ``red nuggets", i.e., nearby massive and small galaxies containing very old ($>$ 10 Gyr) stellar populations.

\begin{figure*}
\centering
    \includegraphics[width=15cm]{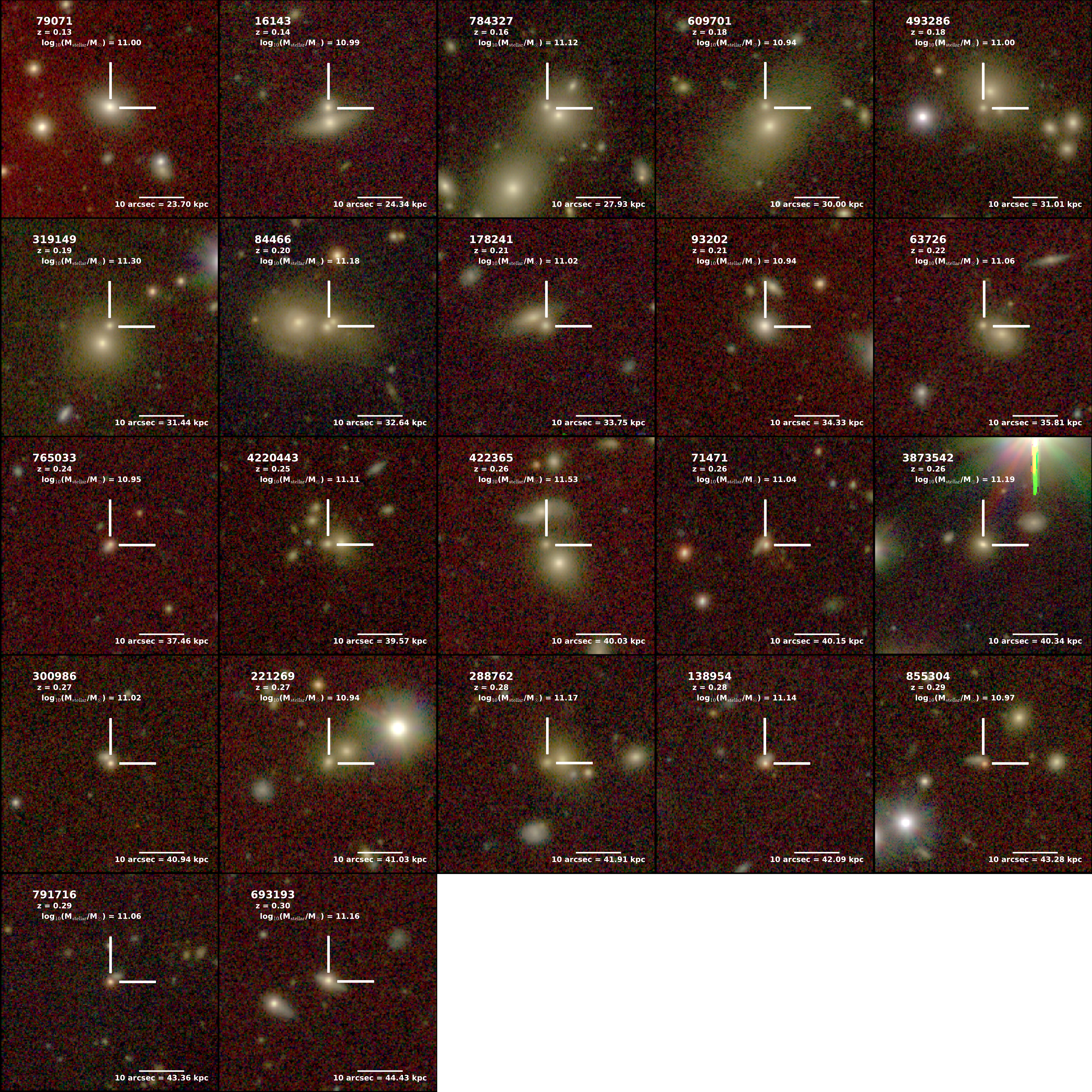}
    \caption{Mosaic with the 50$\times$50 arcsec$^2$ individual images of the 22 galaxies composing our M$_{\star}$ $>$ 8$\times$10$^{10}$M$_{\odot}$ sample. Colour-composite RGB images are created with the $i$-band (red); $r$-band (green) and $g$-band (blue) images from KiDS. The markers pinpoint the exact position of the target galaxy. The surface brightness ranges between 18 and 27\,mag\,arcsec$^{-2}$ in each band. The galaxy name, its spectroscopic redshift and its stellar mass are shown in each image, along with a scale bar with a physical scale.}
    \label{fig:mosaic_rgb}
\end{figure*}

\begin{figure*}
\centering
    \includegraphics[width=15cm]{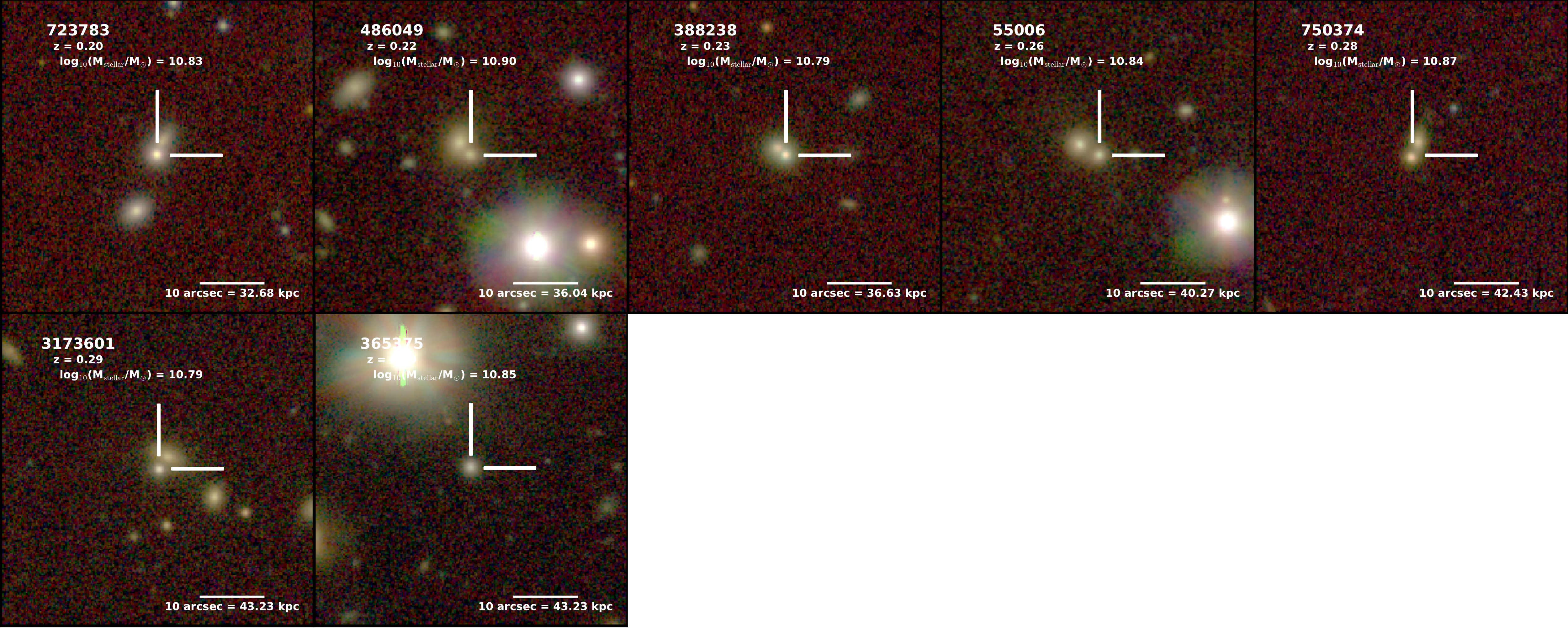}
    \caption{Mosaic with the 50$\times$50 arcsec$^2$ individual images of the 7 galaxies that display effective radii less than 2 kpc in two bands but with stellar masses between 6$\times$10$^{10}$M$_{\odot}$ and 8$\times$10$^{10}$M$_{\odot}$. The RGB stamps are created with the $i$-band (red), $r$-band (green) and $g$-band (blue) images from KiDS. The markers pinpoint the exact position of the target galaxy. The surface brightness levels extend from 18 to 27\,mag\,arcsec$^{-2}$ in each band. The galaxy name, its spectroscopic redshift and its stellar mass are shown in each image, along with a scale bar with the physical scale.}
    \label{fig:mosaic_rgb_borderline}
\end{figure*}

\citet{Poggianti13} \citep[see also][]{Valentinuzzi10a,Valentinuzzi10b} suggested that these surviving compact massive galaxies (the so-called relic galaxies) were to be found in galaxy clusters, where the high velocity dispersion in such overdensities would prevent further merging, as long as the galaxy entered the cluster early enough. This fact was later corroborated by simulations \citep{Quilis13,Stringer15}, although the actual number densities were under debate. The first positive detection of a relic galaxy was NGC1277 \citep{Trujillo14}, which is indeed part of a galaxy overdensity (the Perseus cluster). After this discovery, other papers followed, increasing the number of such extraordinary objects \citep{Stockton14,Ferre-Mateu17,Yildirim17}. Nevertheless,
the actual definition of a relic galaxy is somewhat arbitrary, because of the different compactness criteria, and disagreement regarding the determination of old stellar populations.

Our effort focuses on the selection of a bona-fide sample of nearby Massive Ultracompact Galaxies (MUGs), while detecting relics among them. Massive relic galaxies have exceptional properties. They display both large rotational velocity and velocity dispersion ($\gtrsim$ 300 km/s), show compelling evidence of a bottom-heavy IMF \citep{Martin-Navarro15,Ferre-Mateu17}, host a uni-modal population of globular clusters \citep{Beasley18} and \"UberMassive Black Holes \citep[\"UMBHs;][]{vandenBosch12,Ferre-Mateu15,Yildirim17}. These \"UMBHs are potential outliers in the \citet{Magorrian1998} relation \citep[see][]{Fabian13,Yildirim15,Scharwaechter16}, for a contrasting view, see, e.g. \citet{Emsellem13} or \citet{Graham:16}. It has been argued that this fact supports the view that the gas from these galaxies was stripped when falling into the high-z overdensities, thus accounting for their passive nature and absence of size growth, but to date there is no conclusive observational determination about which environments these massive galaxies inhabit -- there are hints in favour of galaxy overdensities in \citet{Damjanov15} or \citet{Peralta16}, while the opposite could be seen in \citet{Ferre-Mateu17}. 


A number of papers have tried, and succeeded, to identify these objects at lower redshift \citep{Damjanov14,Damjanov15,Saulder15,Tortora16,Charbonnier17,Tortora18}. The difference between our work and theirs is that: 1) by resorting to the Galaxy and Mass Assembly survey \citep[GAMA;][]{Driver11,Baldry18}, we are using the low-redshift survey with the highest spectroscopic completeness to date; 2) With the GAMA data, one could assess the kind of overdensities these objects inhabit -- as we spectroscopically confirm the presence of neighbours (our completeness details could be found in Sections \ref{subsec:criteria} and \ref{subsec:spectral_props}); 3) by having spectra for all members of our target galaxy sample one could obtain both an accurate MUG number density as well as an estimate of their stellar ages (good enough to discern regular MUGs from potential galaxy relics). Moreover, our work introduces subtle differences about the characterization of the structural parameters, the stellar mass calculations and the inclusion or not of photometric redshifts (please also see the end Section \ref{subsec:number_densities} for a detailed enumeration of these caveats).

This paper is structured as follows: Section \ref{sec:data} shows the photometric and spectroscopic data and explains the galaxy selection criteria. Section \ref{sec:results} displays the number of tests applied to determine the structural and stellar population properties of the galaxies within our sample. Section \ref{sec:conclusions} highlights our primary conclusions.  Appendix \ref{app:simus_robustness_struct_params}
shows our simulations to determine the errors in our structural parameter determination. Appendix \ref{app:struct_params_g_r_i} contains tables of our inferred structural parameters. Our assumed cosmology is $\Omega_m$=0.3, $\Omega_\Lambda$=0.7 and H$_0$=70 km s$^{-1}$
 Mpc$^{-1}$. We use a \citet{Chabrier03} Initial Mass Function (IMF), unless otherwise stated. 
Magnitudes are provided in the AB system \citep{Oke1983}.


\section{Data}
\label{sec:data}

\subsection{Galaxy selection criteria}
\label{subsec:criteria}

We retrieve our sample from the GAMA-II database, a panchromatic
galaxy survey providing a set of spectroscopic redshifts down to
$r_{\rm AB}$=19.8\,mag \citep{Liske15}. We focus on the equatorial
fields, that cover 179.94\,deg$^2$ in three separate regions \citep[see Table 1 in][]{Baldry18}, with a high
($\sim$98.5\%) spatially uniform redshift completeness that makes it
optimal for studies of environment \citep[see,
e.g.,][]{Robotham11,Brough13}. We note that in GAMA the same fields were
repeatedly visited, so that, by construction, the spectroscopic
completeness is very high, not only in general, but also over small
scales, avoiding the standard issues found in SDSS spectroscopic data
sets regarding fibre collision.  The tiling and observing strategies
of the survey are discussed in detail in \citet{Robotham10}
and \citet{Driver11}.

Star-galaxy separation is described in \citet{Baldry10}, notably incorporating a selection of marginally-resolved sources by use of near-infrared colours in addition to the usual Sloan Digital Sky Survey \citep[SDSS;][]{Abazajian09} profile separator.

Our preselection starts by compiling the set of
massive galaxies, defined as those with a stellar mass above
8$\times$10$^{10}$M$_\odot$, to be able to compare with previous results in the literature.  
The sample is extracted from version v20 of the catalogue of stellar masses in the GAMA
survey \citep{Taylor11}, and restricted to the
0.02$<$z$<$0.3 redshift range. The adopted redshift is the so-called Z\textunderscore TONRY parameter, which corrects for the Virgo-cluster infall at low redshift and uses the Cosmic Microwave Background frame at z$>$0.03 \citep{Tonry00}.

We note that compact massive galaxies, although already scarce at $z =$2-3 \citep[with comoving number densities $\sim$10$^{-4}$ Mpc$^{-3}$;][]{vanDokkum10,Conselice11,Buitrago13,Barro13}, in the low-z Universe become very rare objects, with comoving number densities
around $10^{-6}$ per Mpc$^3$ \citep{Damjanov14,Damjanov15,Tortora16,Charbonnier17}. It is thus imperative to make sure that
we start with a conservative enough selection criteria, to ensure we do not miss any potential
candidates. Therefore, from the sample of massive galaxies in GAMA, we select those with original 
SDSS-based effective radii R$_{\rm e}$ $<2$\,kpc \citep{Kelvin12} in any of the 
best spatial resolution photometric bands we study  (i.e. \textit{gri}). This cut yields a total of 262 candidates. 
We compile the Kilo-Degree Survey \citep[KiDS;][]{deJong17} and VISTA Kilo-Degree Infrared Galaxy survey \citep[VIKING;][]{Edge13} images of this sample, analyzing them with our bespoke profile fitting code\footnote{Our profile fitting algorithms can be found at \object{{\tt https://github.com/fbuitrago/Profile-fitting}}.} (see Subsection \ref{subsec:str_params}). We visually inspect these galaxies, removing those with GAMA ID (CATAID) 373300 and 537226 as their catalogue-based sizes were only representative of their bulge. Our final sample is defined  by those objects that, in the new analysis, consistently give small effective radii (R$_e<2$\,kpc) in {\sl at least} 2 bands, thus avoiding single-band fluctuations in the size determination. This selection reduces the sample to 34 galaxies. 

Stellar masses were derived according to the methodology defined in \citet{Taylor11}. There, the authors demonstrated that the $(g-i)$ colour is a very good proxy of the stellar mass-to-light ratio $M_{\star}/L_{i}$. Therefore, the best mass results stem from obtaining the galaxy i-band magnitude free from any systematics and then multiplying by its corresponding mass-to-light ratio. In our case, we can estimate more accurately the $i$-band total magnitudes of our MUG sample, as we make use of deep KiDS imaging, in contrast to the SDSS imaging from the original estimates. We  modeled the surface brightness profiles with S\'ersic functions, not restricted to the detection region. We have thus corrected all the GAMA stellar mass estimates with S\'ersic fits (assuming the original mass-to-light ratio) by the formula
\begin{equation*}
\log M_{\rm new} = \log M_{\rm GAMA} + 0.4(m_{\rm i,GAMA} - m_{\rm i,new}),
\end{equation*}
where $M$ denotes stellar mass and $m$ denotes galaxy magnitude. Hence, our detailed light modeling also improves the mass estimates. In general, the change is not significant, although for some galaxies the stellar masses become smaller due to the fact that we remove the light contribution from neighbouring sources. Hereafter, we use these new, more accurate mass estimates.

As a consequence, 5 objects are rejected from our sample. Our final selection comprises 22 galaxies with effective radii smaller than 2\,kpc in at least 2 bands, stellar mass $\geq$8$\times$10$^{10}$M$_\odot$ and redshift 0.02 $< z <$ 0.3. Fig.~\ref{fig:mosaic_rgb} shows a mosaic with the RGB (KiDS) images of our sample of compact massive galaxies. Another 7 galaxies fulfill the size criterion and, given the mass determination uncertainties ($\sim$0.3\,dex), could potentially enter in our sample. Moreover, we note that compact relics feature bottom-heavy IMFs that
could potentially increase the standard IMF-based M/L by a similar amount \citep{Martin-Navarro15}. This additional set appears in Fig. \ref{fig:mosaic_rgb_borderline}, and we will show them separately henceforth. 

\input{table_main_props.tex}
~\\

\subsection{Structural parameter determination}
\label{subsec:str_params}

\begin{figure*}
\centering
    \includegraphics[width=16cm]{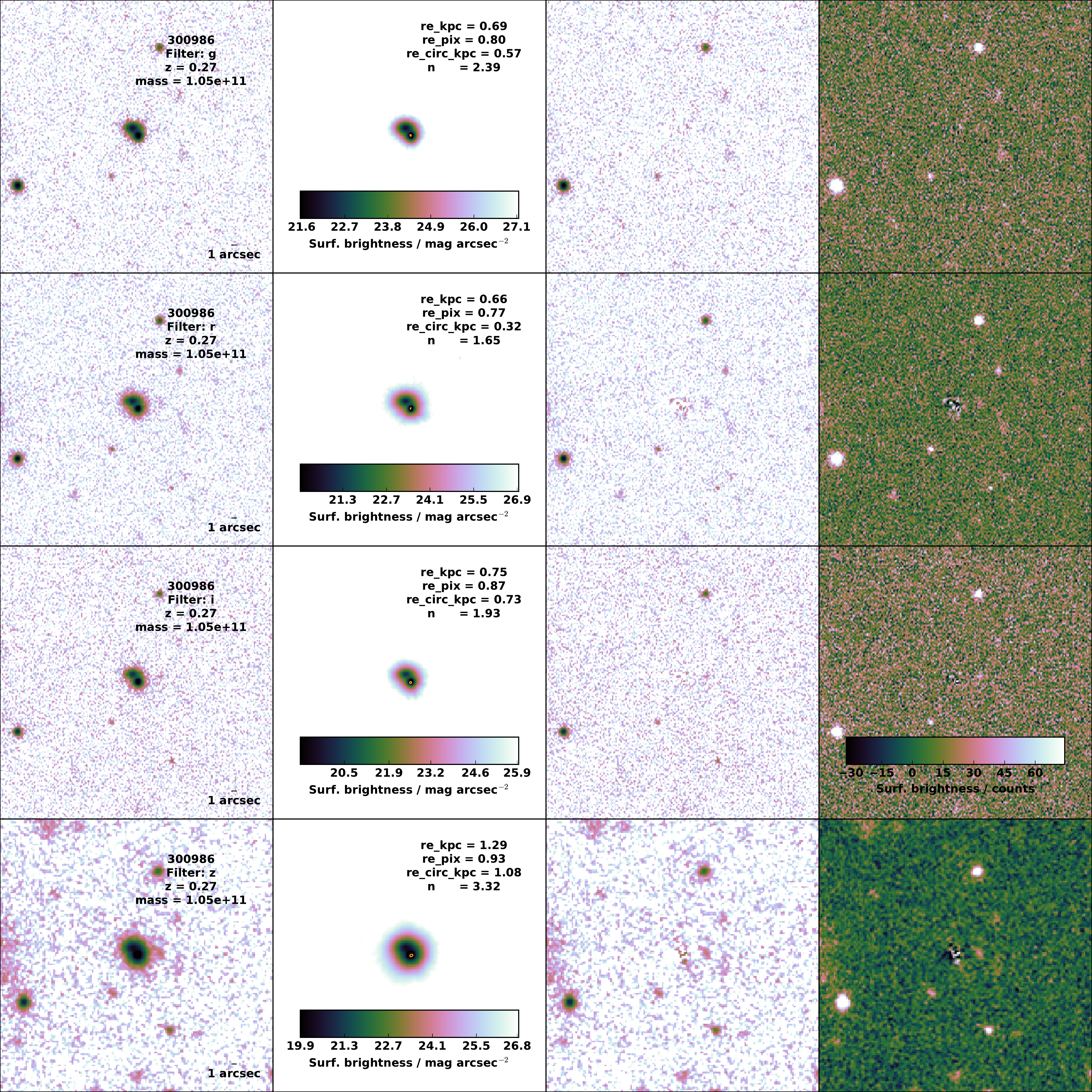}
    \caption{Results from our single S\'ersic fitting analysis for galaxy CATAID 300986. The rows correspond to the photometric bands: (from top to bottom) KiDS $g$, KiDS $r$, KiDS $i$, and VIKING $Z$. The columns are (from left to right) the original 50$\times$50\,arcsec$^2$ image, its GALFIT model with its associated effective radius ellipse, the residual image (original -- model) --in units of surface brightness mag\,arcsec$^{2}$, that is the reason why one colour bar per band-- and the residual image again in units of counts with a single colour bar which is shared by all the images in the last column.}
    \label{fig:example_single_fits}
\end{figure*}

We cut 50 arcsec postage stamp images in the KiDS $g$-, $r$- and $i$-bands and the VIKING $Z$-band. We fit single \citet{Sersic1968} functions to the surface brightness profiles of the galaxies in our photometric images. We applied the same procedures developed in \citet{Buitrago08,Buitrago13}. In short, this pipeline first detects the objects in each image applying {\sc SExtractor} \citep{Bertin1996}, and then uses {\sc GALFIT} \citep{Peng02,Peng10}, a code that convolves the 2D galaxy S\'ersic models with the PSF of the images and determines the best fit by comparing the convolved model with the observed galaxy surface brightness distribution using a Levenberg-Marquardt algorithm to minimize the $\chi^2$ of the fit. The representative PSFs were obtained by the standard procedure of taking isolated non-saturated stars within our imaging and also from the shapelet-based models for weak lensing developed in \citet{Kuijken15}. Neighbouring objects were masked or fitted along with the target galaxy depending on their proximity. At the end of this process, each galaxy was fitted with 4 different PSFs in order to account for the variability of this parameter, taking as our final result the one that displayed the best $\chi^2$ value. An example of our fits is shown in Fig. \ref{fig:example_single_fits}. 

Since we have access to the PSFs utilized by the weak lensing community, we can see how optimal they are to compute galaxy sizes applying our algorithms. The shapelet-based PSF was the chosen one for 3/6/5/3 galaxies in the $g$/$r$/$i$/$Z$ bands (i.e. in 10\%, 21\%, 17\%, 10\% of the cases for our 22+7 object sample). If we only take the shapelet-based PSFs, the sizes deviate on average (3$\sigma$ clipped mean of the quantity (r$_{\rm e,shapelets}$-r$_{\rm e,best~\chi^2}$)/r$_{\rm e,best~\chi^2}$) by $-$25\%/$-$10\%/4\%/$-$30\%. Hence, it seems that the automatic pipelines for weak lensing measurements could be trusted in order to obtain galaxy preselections based on galaxy sizes that will later be refined by the galaxy evolution community.

\begin{figure}
\centering
    \includegraphics[width=8cm]{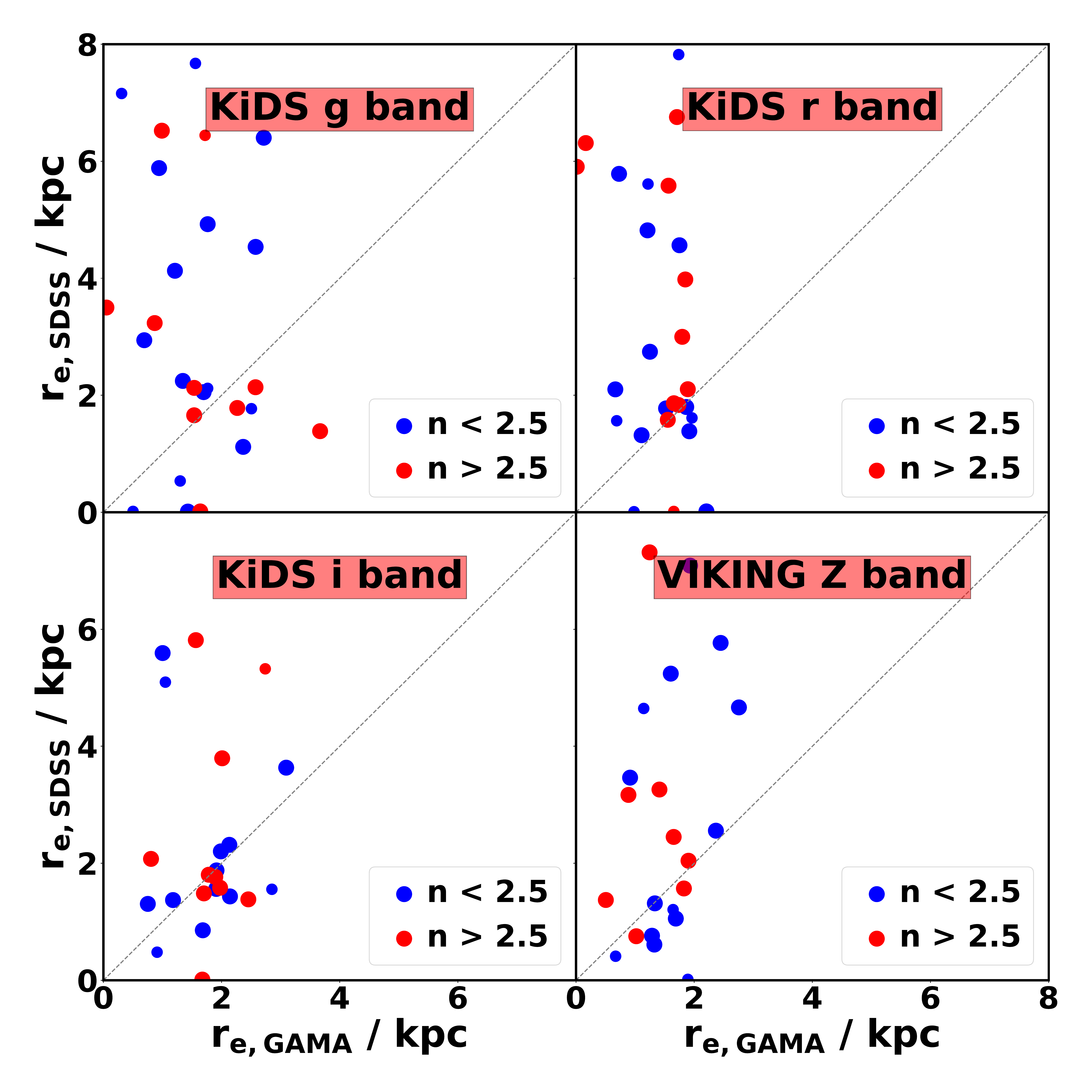}
    \caption{Inferred effective radii in KiDS ($g$-, $r$- and $i$-bands) and VIKING imaging ($Z$-band) versus the effective radii in SDSS \citep[as described in][]{Kelvin12} for the 22+7 objects in our sample. The colour coding follows the S\'ersic index, with blue dots being disk-like ($n <$ 2.5) objects and red dots being spheroid-like ($n >$ 2.5) objects. The symbol sizes indicate stellar mass, with big symbols representing $>$ 8$\times$10$^{10}$M$_{\odot}$ galaxies, and small symbols between 6$\times$10$^{10}$M$_{\odot}$ and 8$\times$10$^{10}$M$_{\odot}$. There is little correlation between sizes in the two axes, perhaps only for small galaxies in SDSS being also small in the deeper imaging.}
    \label{fig:mass_size_rels}
\end{figure}

We conducted several tests to prove the robustness of our structural parameters. In the first instance, we simulated, using again {\sc GALFIT}, 2D S\'ersic functions convolved with the PSFs of the images and injected them into the science images. We add a full description of this procedure in Appendix \ref{app:simus_robustness_struct_params}. In brief, the results produce realistic errors for the structural parameters and tell us that caution must be applied to results derived from faint and concentrated (high S\'ersic index) objects. Additionally, we compare in Fig. \ref{fig:mass_size_rels} the effective radii derived from the deep KiDS and VIKING imaging with previous results for the same galaxies, based on the shallower SDSS data \citep{Kelvin12}. It is clear that there is little (if any) correlation between these size estimates, with SDSS featuring a wider range of effective radii. This result is expected as the SDSS images are shallower ($\sim$2\,mag with respect to KiDS; $\sim$1.5\,mag relative to VIKING), and have a coarser sampling  (0.396 arcsec/pix in SDSS versus 0.21/0.339\,arcsec/pix in KiDS and VIKING respectively), implying uncertain effective radii measurements particularly for such small galaxies.

\begin{figure*}
\centering
    \includegraphics[width=14cm]{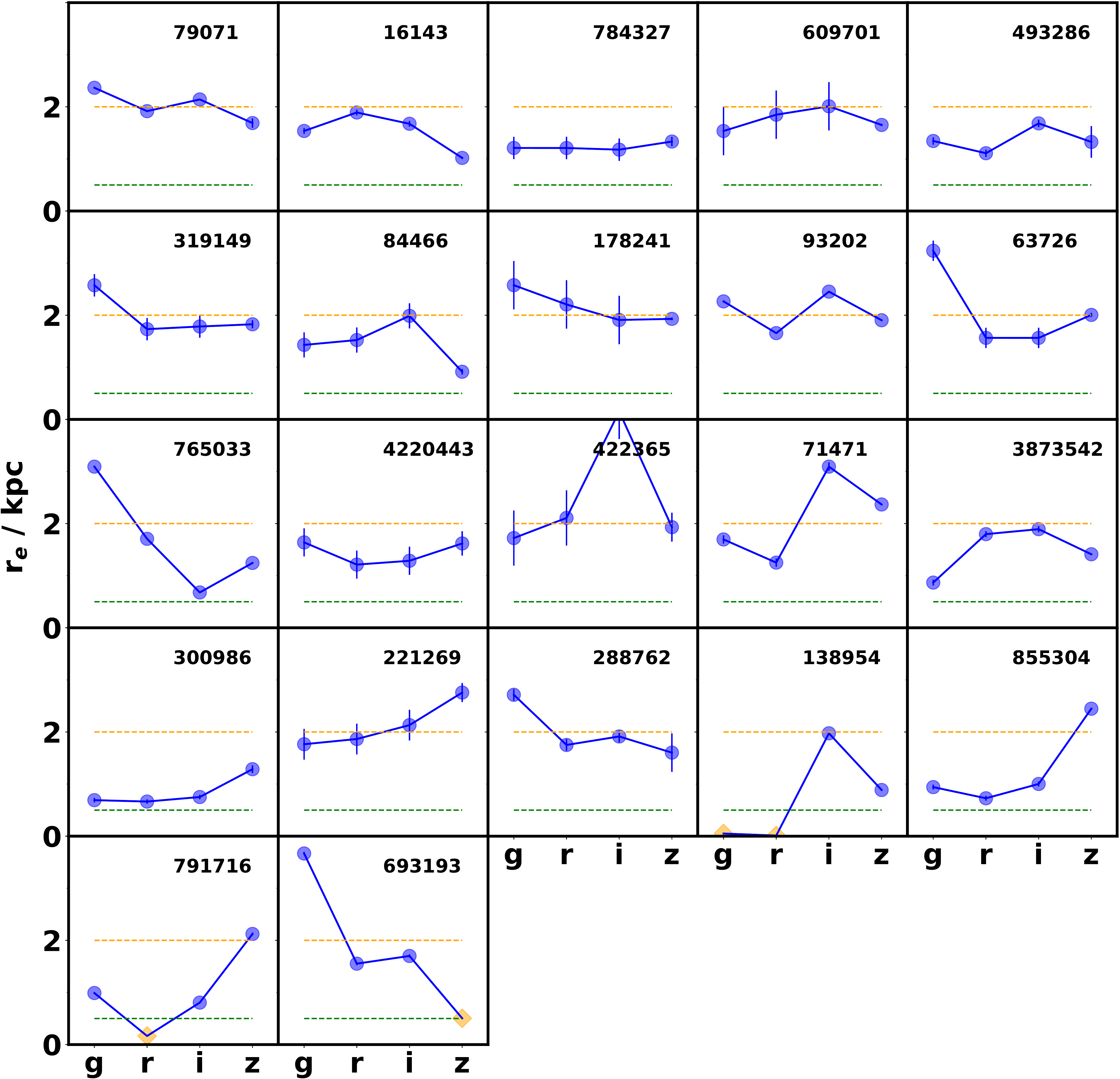}
    \caption{Mosaic displaying the effective radii in each band for the 22 galaxies in our MUG sample. As reference, the horizontal dotted lines correspond to R$_{\rm e}$ = 0.5\,kpc and R$_{\rm e}$ = 2\,kpc. Bad fits are denoted by orange diamonds.}
    \label{fig:mosaic_sizes_good}
\end{figure*}
\begin{figure*}
\centering
    \includegraphics[width=14cm]{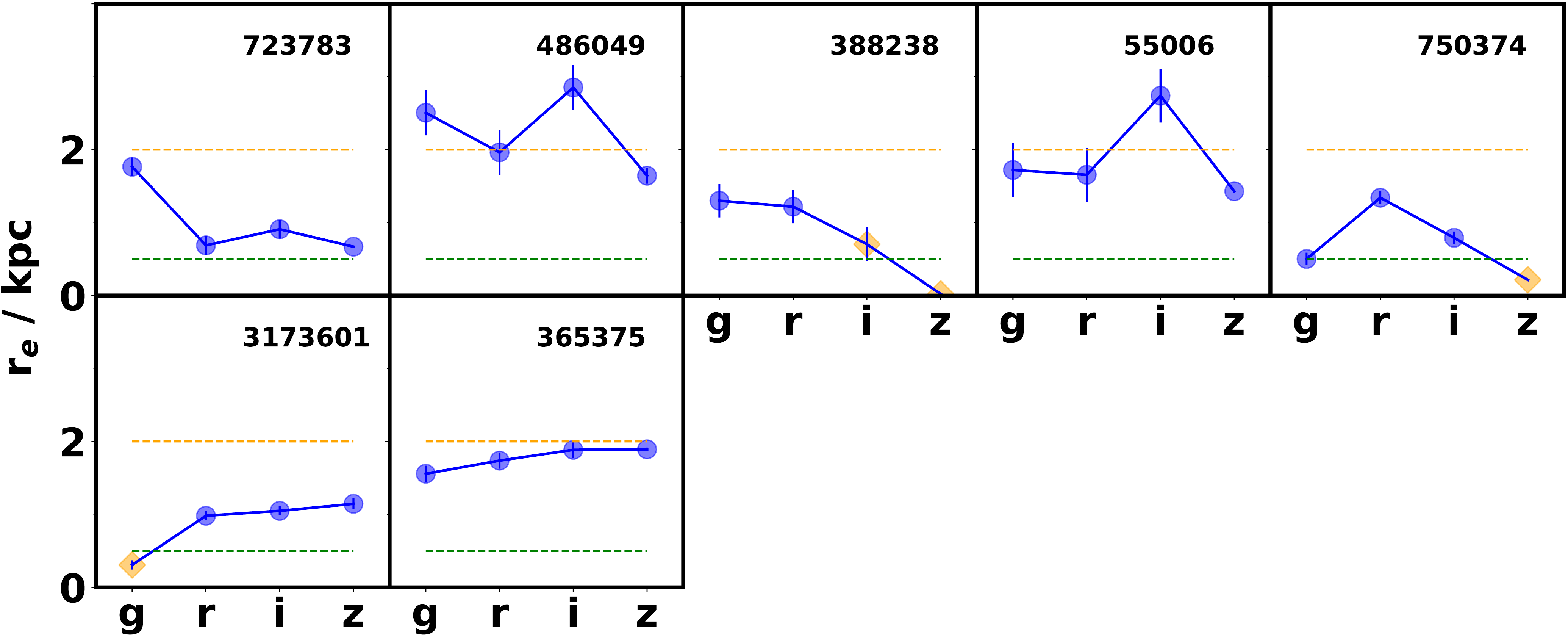}
    \caption{Mosaic displaying the effective radii in each band from the 7 galaxies that display effective radii less than 2 kpc in two bands but with stellar masses between 6$\times$10$^{10}$M$_{\odot}$ and 8$\times$10$^{10}$M$_{\odot}$. As reference, horizontal dotted lines correspond to R$_{\rm e}$ = 0.5 kpc and R$_{\rm e}$ = 2 kpc. Bad fits are denoted by orange diamonds.}
    \label{fig:mosaic_sizes_borderline}
\end{figure*}

The structural parameters of our sample are shown in Table \ref{tab:struct_param_r_band} ($r$-band, best seeing band), and also in Appendix \ref{app:struct_params_g_r_i}: Tables \ref{tab:struct_param_g_band}.1 for the $g$-band, \ref{tab:struct_param_i_band}.2 for the $i$-band and \ref{tab:struct_param_z_band}.3 for the $Z$-band. Objects that produced non-valid fits, i.e. their fits are in the constraints of our analysis, are excluded from both the tables and the plots. The effective radii are also shown graphically in Fig. \ref{fig:mosaic_sizes_good} for the 22 object sample and in Fig. \ref{fig:mosaic_sizes_borderline} for the extra 7 objects. Interestingly, these plots show that for most of the galaxies, there is little variation in the effective radius for all bands, with some exceptions preferentially in the bluest band where the light distribution could be slightly patchier \citep{Buitrago08}. 

\input{struct_param_table_r_band.tex}
~\\

\section{Results}
\label{sec:results}

\subsection{The size-mass relation}
\label{subsec:size-mass}

Central to the problem of the evolution of massive galaxies is the quantification of the compactness of their stellar component. There has been a range of potential definitions. Several key factors must be taken into account:
\begin{itemize}
   \item Threshold in mass, size and/or stellar density;
   \item Photometric band (i.e. rest-frame wavelength) used in the derivation of galaxy sizes;
   \item Depth of the observations;
   \item How the sizes are inferred (S\'ersic function effective radius, {\sc SExtractor}-derived half light radius, non parametric fit radius, number and assumed values for the structural parameter components, e.g. fixing or not their S\'ersic index);
   \item Use of circularized or semi-major axis ``radii''.
\end{itemize}

\begin{figure*}
\centering
    \includegraphics[width=17cm]{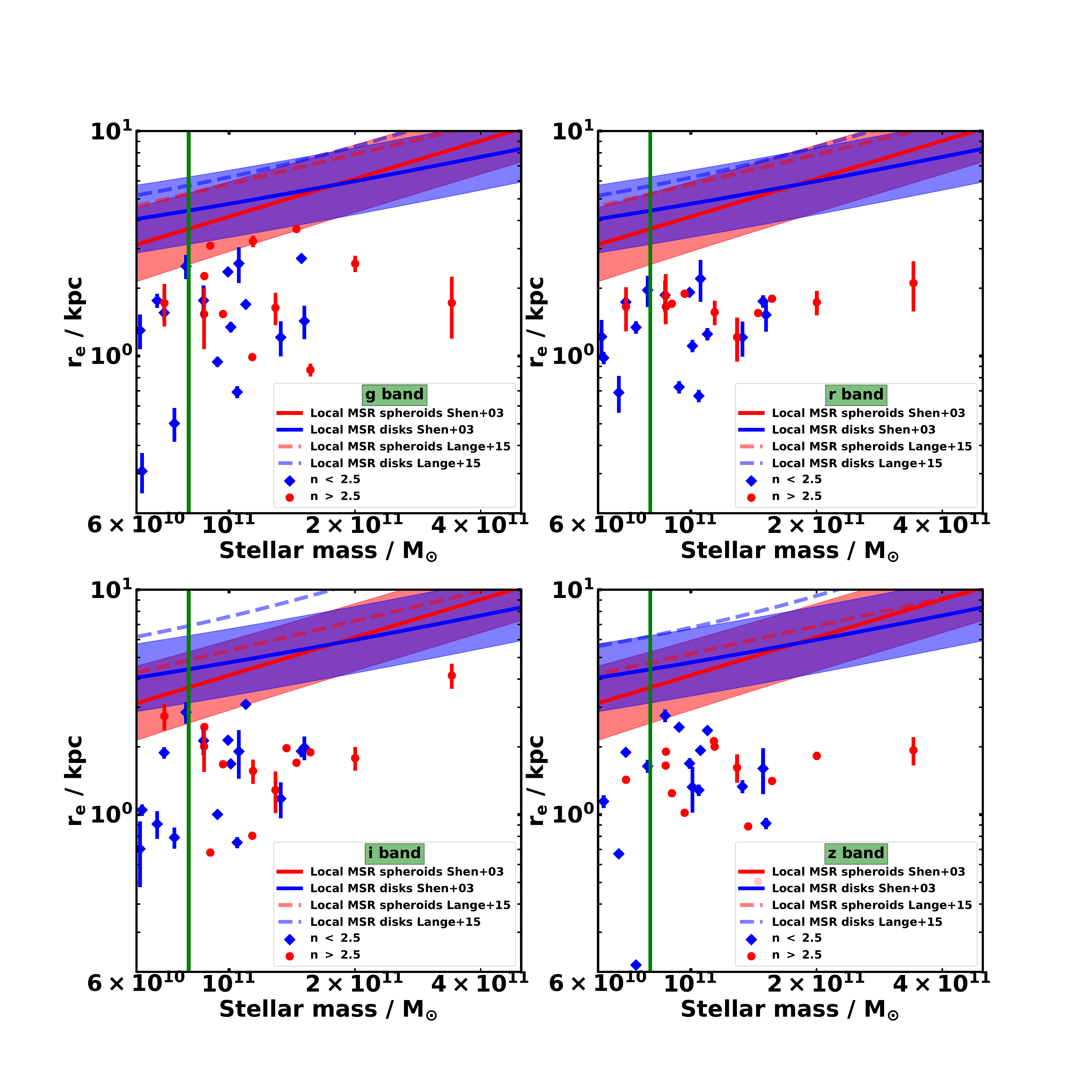}
    \caption{Mass-size relation in all bands included in our study (from top to bottom, from left to right $g$-, $r$-, $i$- and $Z$-band. Objects are split with respect to the surface brightness profile into disk-like ($n <$ 2.5; in blue) and spheroid-like ($n >$ 2.5; in red). The green vertical line separates the objects in the main sample (M$_{\rm stellar}$  $\geq$8$\times$10$^{10}$M$_\odot$) from those that might be compatible if we include the mass uncertainties (M$_{\rm stellar}$  $\geq$6$\times$10$^{10}$M$_\odot$). The solid regions represent the canonical SDSS $z$-band mass-size relation from \citet{Shen03}, including the dispersion. The dashed lines are the equivalent trends for GAMA galaxies from \citet{Lange15}, for each photometric band. The error bars are derived according to the results in our simulations. Note the effective radii for our sample are not circularized.}
    \label{fig:mass_size}
\end{figure*}

Fig. \ref{fig:mass_size} displays the mass-size relation of our sample for all the bands at study, split with respect to morphology, between disk- and spheroid-like (S\'ersic index $n <$ 2.5 and $n >$ 2.5, respectively). The local SDSS size-mass relations \citep[][and their scatter]{Shen03} are overplotted, derived from circularized $z$-band effective radii. If we correct these relations to match our semi-major axis measurements (i.e. non-circularized), they will shift slightly to higher values, matching closer the trends shown with dashed lines, that correspond to the relation inferred for GAMA \citep{Lange15}. The GAMA relations we show are the ones corresponding to each photometric band, with the same analytical parametrizations for disk- and spheroid-like objects as those in \citet{Shen03}.

It is clear that the objects in our sample feature smaller sizes (at the 2-3\,$\sigma$ level) than those expected in galaxies with the same stellar mass in the local relation. Given the large area covered by the observations, we cannot quote a reference value of the PSF FWHM, but the KiDS documentation states that the $r$-band (6231 \AA\, pivot wavelength) data features the smallest seeing ($\leq$0.7\,arcsec). Therefore, we will take the $r$-band as the reference filter throughout this paper (rest-frame $\sim$5500-5000\AA\, for the galaxies in our sample). The $r$-band mass-size relation features the lowest scatter and smallest sizes, reassuring us in the compactness of our galaxy sample. Conversely, we could attribute at least part of the scatter of the galaxy sizes in other bands to the worse seeing conditions, specially taking into account the small size variations expected in massive galaxies, which are usually quiescent objects at z$<$0.3 \citep{Cassata10,Cassata11}.

\subsection{On the S\'ersic index values for our sample}
\label{subsec:sersics}

\begin{figure}
\centering
    \resizebox{\hsize}{!}{\includegraphics{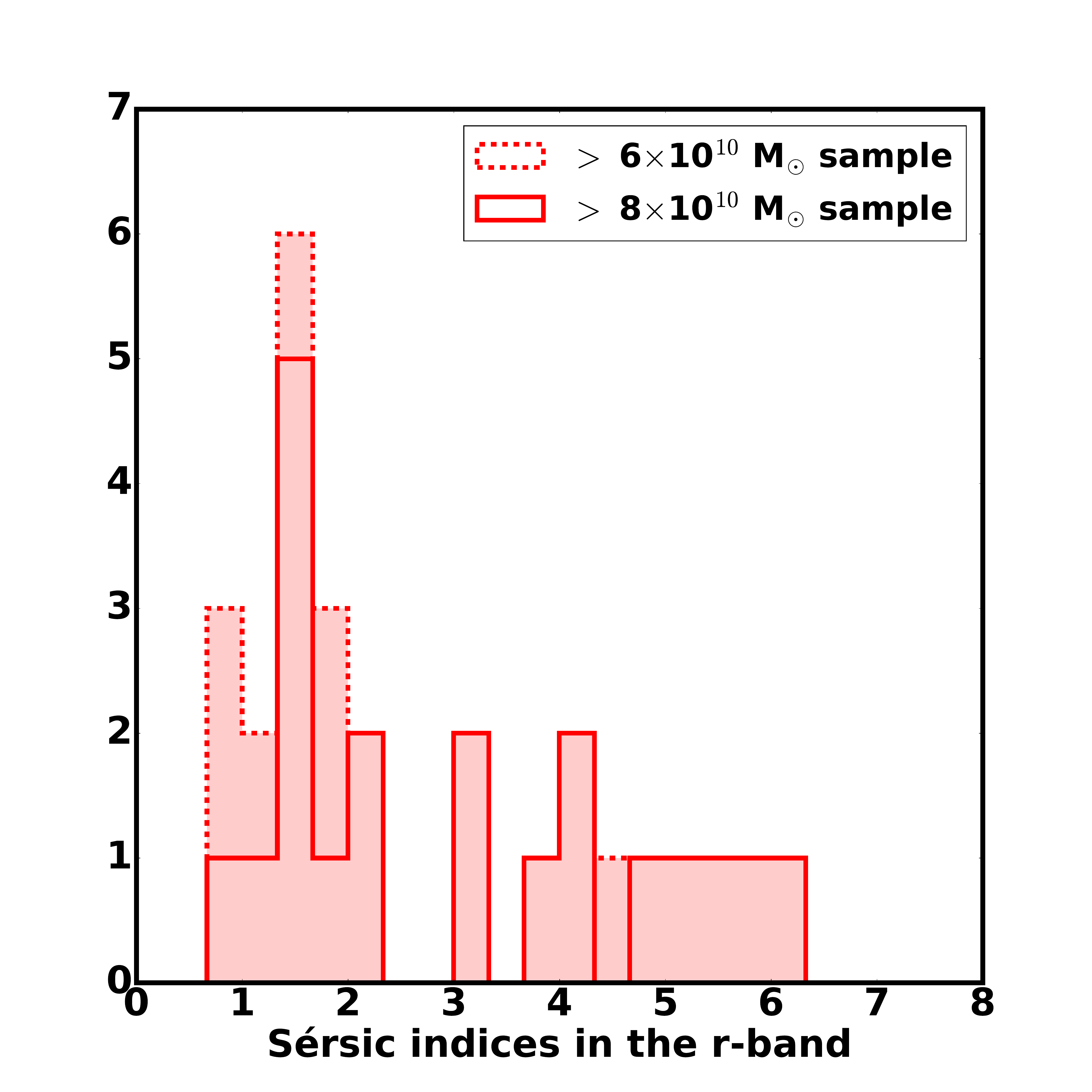}}
    \caption{Histogram of the S\'ersic indices in the $r$-band -- our filter with the best seeing ($\leq$0.7\,arcsec). The solid histogram corresponds to MUGs with M$_{\star}$ $>$ 8$\times$10$^{10}$ M$_{\odot}$ (note that only 20 objects appear as the other 2 have non-valid analyses in the $r$-band) while the dashed ones are for 6$\times$10$^{10}$ $<$ M$_{\star}$/M$_{\odot}$ $<$ 8$\times$10$^{10}$ galaxies. This histogram reveals that the majority of the objects are disk-like ($n <$ 2.5), in accordance to the ``swollen disk'' morphology found in relic galaxies and in general for MUGs at low- and high-z.}
    \label{fig:sersic_r}
\end{figure}

Fig.~\ref{fig:sersic_r} displays the S\'ersic index values in the $r$-band for the 22+7 MUGs in our sample. This morphological information is especially valuable at the redshifts covered by our sample. At lower redshift, the galaxy components and traits usually allow an easy visual characterization, while the small (physical and apparent) sizes of the target galaxies -- and hence very few pixels subtended in the detector -- hinder a robust visual classification of morphology. Half of the 22 objects in the main sample are disk-like and the other half are spheroid-like, although there is a peak at $n=$1.5-2, corresponding to the early disk morphologies seen both at low- and high-z by the Hubble Space Telescope \citep{vanderWel11,Buitrago13,Trujillo14,Ferre-Mateu17} and also with ground-based Adaptive Optics observations \citep{Carrasco10,Trujillo12a,Stockton14}. Interestingly, note that the additional set of 7 lower mass galaxies only features one spheroid-like galaxy (as measured in the $r$-band).

\subsection{2D and 3D stellar densities}
\label{subsec:densities}

\begin{figure*}
\centering
    \resizebox{\hsize}{!}{\includegraphics{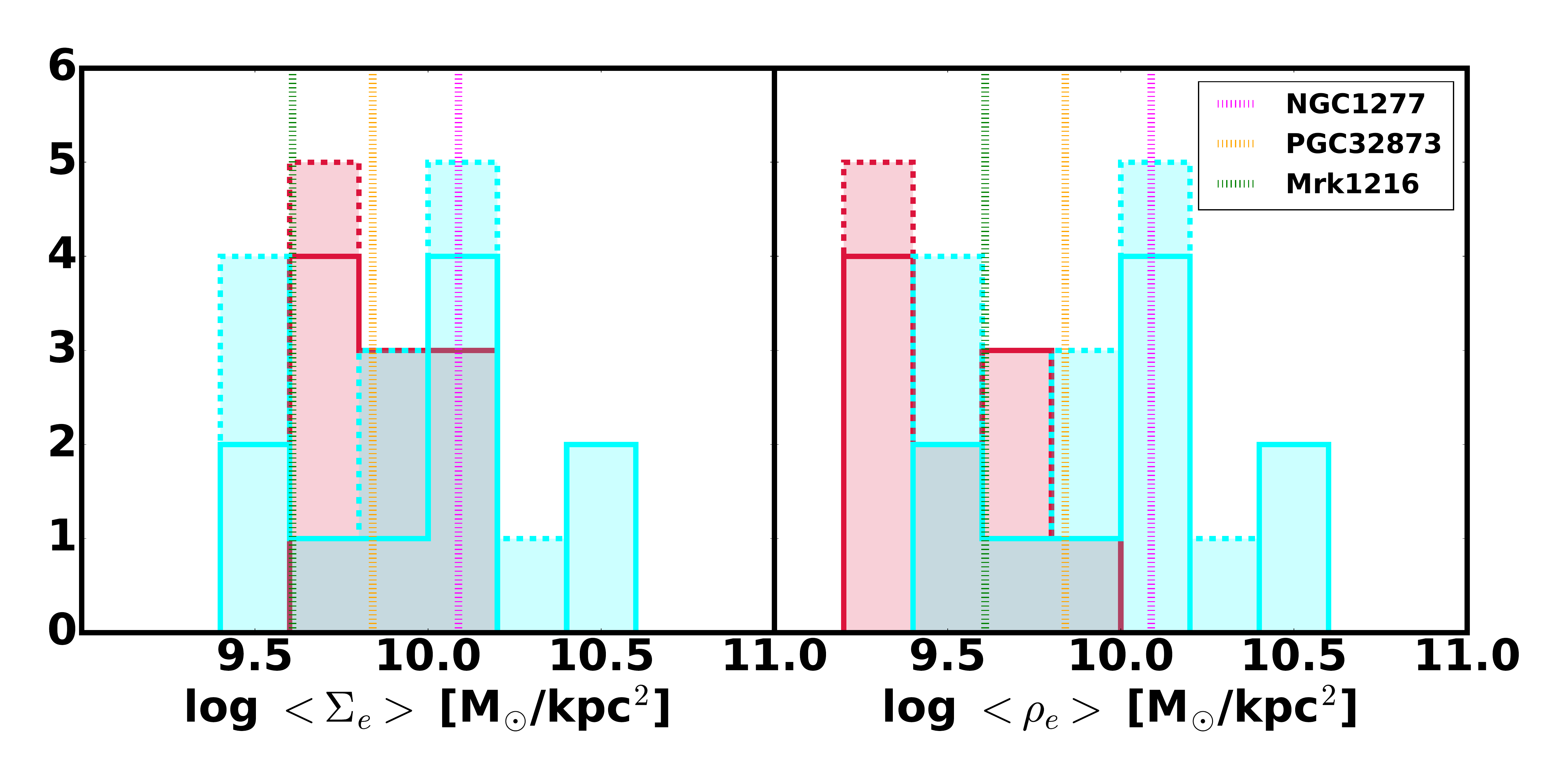}}
    \caption{\textit{Left: } Histogram with the 2D stellar mass density within one effective radius ($<\Sigma_{\rm e}>$ = M$_{\star}$/2$\pi$R$_{\rm e}^{2}$) of our MUG sample, split between disk- (blue) and spheroid-like (red) morphology ($n <$ 2.5 and $n >$ 2.5, respectively). The solid histogram denotes the objects with M$_{\star}$ $>$ 8$\times$10$^{10}$ M$_{\odot}$ and the dashed one those with  6$\times$10$^{10}$ $<$ M$_{\star}$/M$_{\odot}$ $<$ 8$\times$10$^{10}$. Overplotted, as vertical dotted lines, are the values for the confirmed relic galaxies in \citet{Ferre-Mateu17}. The MUGs in our study span the range of values expected of relic galaxies. \textit{Right: } Histogram showing the 3D stellar mass density within one effective radius of our MUG sample, again dividing it between M$_{\star}$ $>$ 8$\times$10$^{10}$ M$_{\odot}$ (solid histogram) and 6$\times$10$^{10}$ $<$ M$_{\star}$/M$_{\odot}$ $<$ 8$\times$10$^{10}$ (dashed histogram). We split the galaxies into disk- (blue) and spheroid-like (red) morphology ($n <$2.5 and $n >$2.5, respectively). We also compute their densities differently: $<\rho_{\rm e}>$=M$_{\rm stellar}$/2$\pi$R$_{\rm e}^{2}$h for disks (with $h =$ 1 kpc), and $<\rho_{\rm e}>$=$3$M$_{\rm stellar}/8\pi R_{\rm e}^{3}$ for spheroids. Overplotted are the values of the confirmed local relic galaxies from \citet{Ferre-Mateu17} -- using our formula for disks. On average, disks appear slightly denser than spheroids.}
    \label{fig:hist_densities_in_2D}
\end{figure*}

We show the stellar densities for the objects in our main sample in Fig. \ref{fig:hist_densities_in_2D} derived from the effective radii  in the $r$-band, as this is the photometric band with the best seeing ($\leq$0.7\,arcsec). The solid histogram represents the objects with M$_{\star}$ $>$ 8$\times$10$^{10}$ M$_{\odot}$ and the dashed one corresponds to those with  6$\times$10$^{10}$ $<$ M$_{\star}$/M$_{\odot}$ $<$ 8$\times$10$^{10}$. The left panel in our plot  shows the 2D densities or surface densities within the effective radius and thus the $\langle\rangle$ notation ($\langle\Sigma_{\rm e}\rangle$ = M$_{\star}$/2$\pi$R$_{\rm e}^{2}$). We split our sample between disk-like (blue histogram) and spheroid like objects (red histogram).

We compared our derived densities with those corresponding to the three unambiguously detected relic galaxies (at a distance $\lesssim$106\,Mpc away from the Milky Way) from \citet{Ferre-Mateu17}. There, the authors established a ``degree of relic'', based on the density of the objects and the ages of their stellar populations. The most extreme case is NGC1277, while the galaxy that appears closer to normal is Mrk1216. In Fig.~\ref{fig:hist_densities_in_2D}, we see that the MUGs in our sample span densities related to all kinds of ``degree of relic''. There is also a hint towards higher stellar densities in the disk-like subsample. We attempted to further investigate this aspect by performing the Anderson-Darling test \citep[][a nonparametric two-sample statistic, better suited than Kolmogorov-Smirnov for small samples]{AD52}. However, the statistical significance to reject the hypothesis that disk- and spheroid-like objects come from the same distribution is small (75\% if taking the objects with M$_{\star}$ $>$ 8$\times$10$^{10}$ M$_{\odot}$, 66\% for those with M$_{\star}$ $>$ 6$\times$10$^{10}$ M$_{\odot}$).

To shed more light upon this issue, we show 3D stellar densities within the effective radius for our objects in the right panel of 
Fig.~\ref{fig:hist_densities_in_2D}, also displaying with different colors disk-like ($\langle\rho_{\rm e}\rangle$=M$_{\rm stellar}$/2$\pi$R$_{\rm e}^{2}$h) or spheroid-like ($\langle\rho_{\rm e}\rangle$=3M$_{\rm stellar}$/ $8\pi$R$_{\rm e}^{3}$) morphology. We did similarly for the three local relics from the literature. In all cases (our sample and the three local galaxies) we assume a disk scale-length of h$=$1\,kpc. The Anderson-Darling test now suggests a cleaner separation, rejecting the null hypothesis that disk- and spheroid-like objects come from the same distribution at a 99.5\% significance level for galaxies with M$_{\star}$ $>$ 8$\times$10$^{10}$ M$_{\odot}$ and at a 99.9\% significance level for galaxies with M$_{\star}$ $>$ 6$\times$10$^{10}$ M$_{\odot}$. Nevertheless, these are only tentative results, as we remind the reader about our small number statistics (only 20 objects) and the strong assumptions regarding the 3D densities (i.e. assuming either a perfect disk -- no bulge, no arms, no halo -- with a fixed scale length or a homogeneous sphere without any triaxiality). One could attempt to improve this situation in the future using a dynamical analysis (e.g. Schwarzschild modeling) of 3D spectroscopic observations. 

\begin{figure}
\centering
    \resizebox{\hsize}{!}{\includegraphics{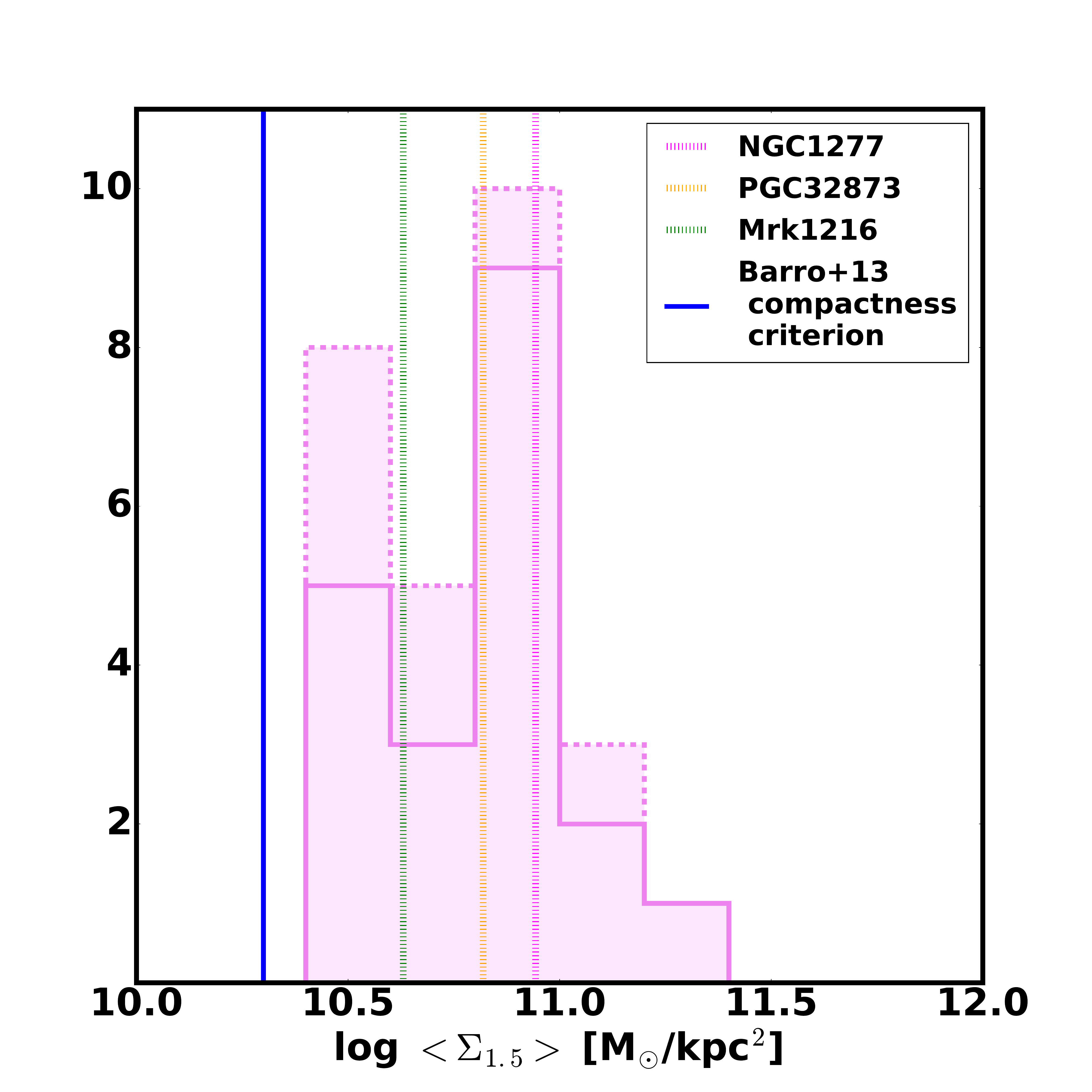}}
    \caption{Distribution of the 2D stellar mass density of our MUGs, following the parameter $\langle\Sigma_{1.5}\rangle\equiv$M$_{\star}$/R$_{\rm e}^{1.5}$ to define compactness, as in \citet{Barro13}. Overplotted are the values for the confirmed nearby relic galaxies given in \citet{Ferre-Mateu17}. The MUGs in our study span a similar  range of density as in relic galaxies, and some even surpass the largest local value of the surface mass density (NGC1277).}
    \label{fig:hist_Sigma_Barro}
\end{figure}

Finally, in Fig.~\ref{fig:hist_Sigma_Barro} we show the compactness criterion following the prescription of \citet{Barro13}, that define a density parameter $\langle\Sigma_{1.5}\rangle$ = M$_{\star}$/R$_{\rm e}^{1.5}$. All the MUGs in our sample inhabit the region delimited by 
$\log\langle\Sigma_{1.5}\rangle$/M$_{\odot}$ kpc$^{-1.5}$ $>$ 10.3 (thick blue vertical line), confirming the uniqueness of our sample.

\subsection{Stellar population content}
\label{subsec:spectral_props}
We use the available optical spectroscopy from GAMA to constrain the
underlying stellar populations of our sample of massive and compact
galaxies.  Only one out of the 22+7 galaxies have an SDSS spectrum (79071). 
We note that an additional set of 9 MUGs 
apparently had a nearby SDSS spectrum, but, on inspection, corresponded 
to a nearby massive galaxy, as many of our galaxies were found to lie
in close pairs or small groups (see Fig.~\ref{fig:mosaic_rgb}). This issue is worth mentioning in the light of the expected incompleteness of the SDSS spectroscopic survey. Another galaxy (93202) only has a low SNR spectrum from the 2dFGRS \citep{2dFGRS} and cannot be used for the analysis of populations. The remaining 20 have spectra from the AAT/AAOmega spectrograph, as well as the additional subsample of 7 potential MUGs, although from this  subsample, galaxy 3173601 has a nQ=2 quality flag, meaning the 
probability of a correct redshift is only 80\%.

Unfortunately, the data have a rather low S/N for
line strength work (the average S/N in the 5000-5500\AA\ rest-frame
window is $\sim$10 per \AA), so
we decided to focus on a targeted set of line strengths, and to
compare with SSP-equivalent parameters.  The spectra are corrected for
foreground extinction, using the colour excess maps of \citet{Schlafly:11},
following a standard dust extinction law for the Milky Way
\citep{Cardelli1989}, and brought to the rest-frame.

\begin{figure*}
\centering
    \resizebox{\hsize}{!}{\includegraphics{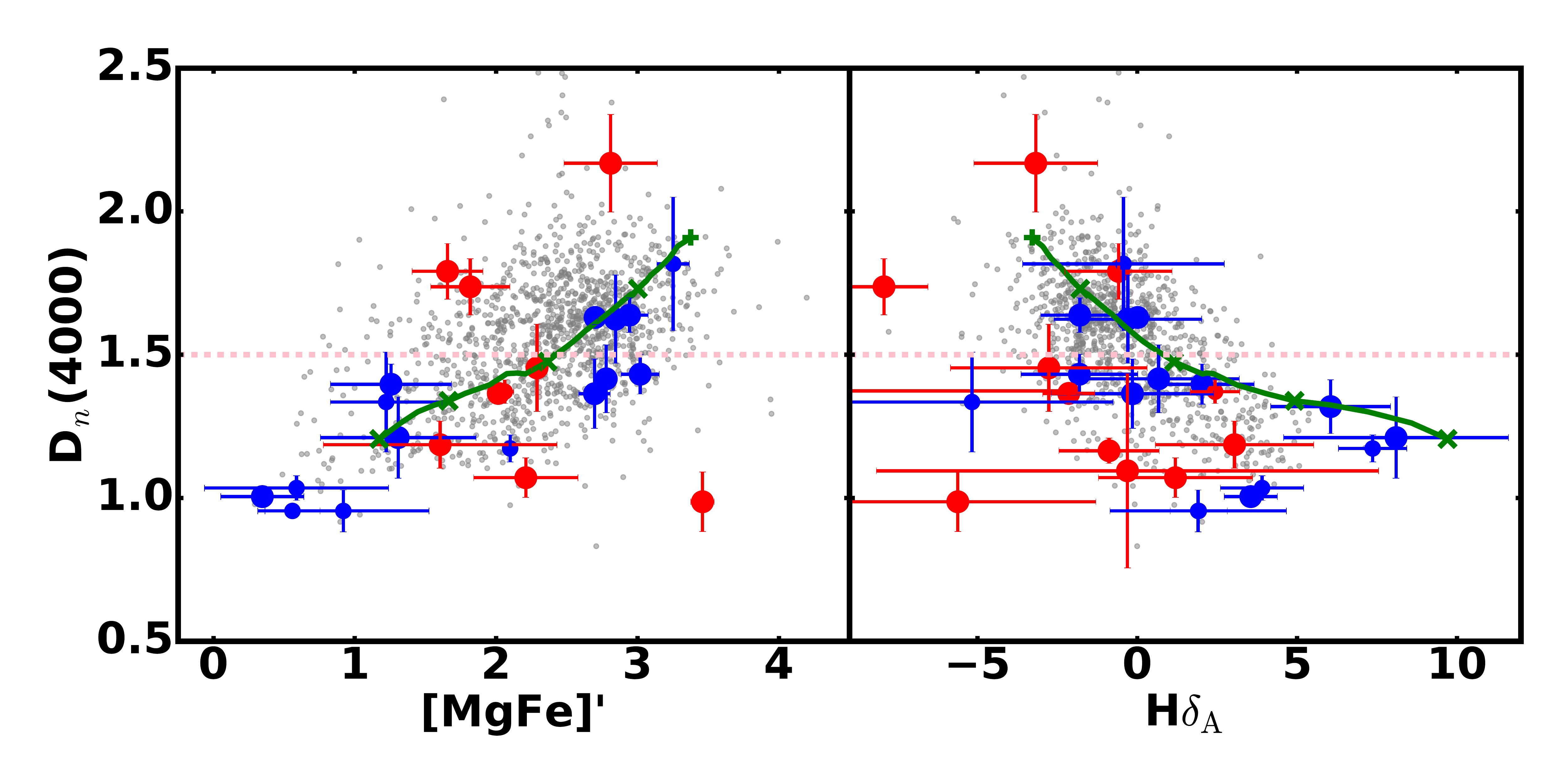}}
    \caption{Distribution of age- and metallicity-sensitive line
      strengths of our MUG sample. The error
      bars are shown at the 1$\sigma$ level. The blue (red) circles
      show the results for galaxies with a S\'ersic index n$<$2.5
      ($>$2.5), with their sizes referring to their stellar masses (6-8$\times$10$^{10}$M$_{\odot}$ are depicted by small circles while $>$ 8$\times$10$^{10}$M$_{\odot}$ are the big circles).  For reference, we include as grey dots the measurements
      of the general sample of GAMA massive galaxies
      ($>10^{11}$M$_\odot$) with AAT/AAOmega spectra. 
      The horizontal
      dashed line -- at D$_n$(4000)=1.5 -- splits the sample between
      old (above the line) and young populations. The green lines
      trace the evolution of a simple stellar population at solar
      metallicity (see text for details).}
    \label{fig:Dn4K}
\end{figure*}

Fig.~\ref{fig:Dn4K} shows the
4000\AA\ break strength \citep[using the D$_{n}$(4000) index
  of][]{Balogh1999}, the Balmer index H$\delta_A$ \citep{Worthey1997}, and
the metallicity-sensitive index [MgFe]$^\prime$ of \citet{Thomas03}.
The error bars are given at the 1\,$\sigma$
level. For reference, we consider D$_n$(4000)=1.5 (dashed horizontal
line) as a threshold to split the sample between ``young'' and ``old''
populations. That value of the index corresponds to an SSP-equivalent
age of 2\,Gyr at solar metallicity. The green lines trace the
evolution of the MIUSCAT models \citep{Vazdekis12} for a set
of simple stellar populations with a Kroupa IMF and solar metallicity,
with ages from 0.5 to 10\,Gyr. The 'x' symbols mark the SSP ages
0.5, 1, 2 and 5\,Gyr, and the '+' symbols mark the oldest age
in the track: 10\,Gyr. As reference, we include in the figure the
results for the sample of massive (stellar mass $>10^{11}$M$_\odot$)
galaxies from the GAMA survey, with AAT/AAOmega spectra (grey dots).

In the figure, the size of the symbols split the sample between the 
bona-fide set (large dots) and the additional 7, lower-mass candidates
(smaller dots). The dots are also colour-coded -- blue for disk-like ($n<$2.5) 
objects and red for spheroid-like systems ($n>$2.5). Although the S/N is not high enough for 
a detailed analysis of the populations, our sample features a wide range of ages, 
with 7 galaxies out of 29 ($\sim$25\%) having older populations (therefore 
'relics'), where both D$_n$(4000) and H$\delta_A$ suggest old ages. These relics are split into four disk-like systems (79071, 221269, 84466 and 486049) and 3 spheroids (3873542, 319149 and 609701). Out of the four disks, 
only one galaxy lacks an indication of a galactic companion (79071), and
also one of the spheroids (3873542) appears as an isolated galaxy. 
We will explore the environment of our MUGs in more
detail in Section \ref{subsec:environment}. The galaxy with the strongest 4000\AA\  break (319149) is expected to have a very high formation redshift (z$_{\rm FOR}\gtrsim$3). It resembles 
the local relic NGC1277: it is a very massive (M$_{\star}=$ 2$\times$10$^{11}$M$_{\odot}$) satellite of an even 
more massive galaxy. Apart from these aspects, this plot shows no clear trend between galaxy 
morphology and mass with respect to the spectral features.

\begin{figure}
\centering
    \resizebox{\hsize}{!}{\includegraphics{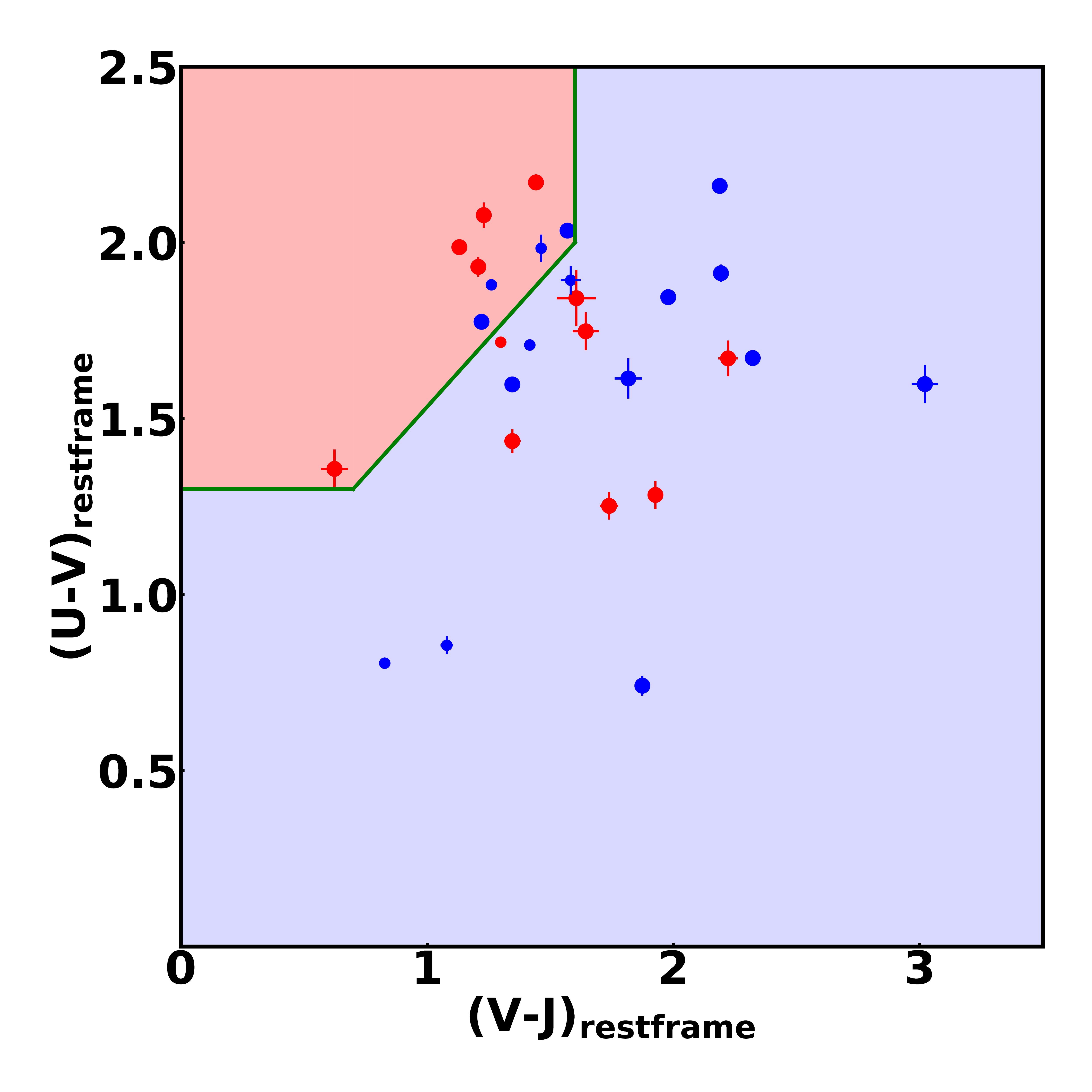}}
    \caption{Rest-frame UVJ diagram \citep[as defined in][]{Williams09} for our MUG sample. The color and size coding is the same as for the rest of the figures: blue and red points distinguish disk-like and spheroid like objects ($n <$ 2.5 and $n >$2.5, respectively), while small and big points denote objects with small and large stellar masses (6-8$\times$10$^{10}$ and $>$ 8$\times$10$^{10}$M$_{\odot}$). The red and blue coloured regions separate quiescent from star-forming galaxies.}
    \label{fig:uvj}
\end{figure}

Fig.~\ref{fig:uvj} shows the UVJ diagram, with divisions as defined in \citet{Williams09}, which has proved to be a powerful tool to photometrically discern quiescent (red region) from dusty star forming (blue region) galaxies \citep[see, e.g.,][]{LADG_UVJ}. Our sample is shown with the same colour- and size-coding as in Fig.~\ref{fig:Dn4K}. The rest-frame colours are derived from comparisons of the
observed D$_n$(4000), H$\delta_A$ and [MgFe]$^\prime$ line strengths with a grid of simple stellar populations
from the models of \citet{BC03}, covering a range of ages (from 0.5 to 13\,Gyr) and metallicities (from [M/H]=$-0.3$ to $0.3$\,dex). The K-corrections were derived from the spectroscopic information, by comparing the observed line strengths with a set of stellar populations, as above. The best fit spectrum is then used to translate 
between rest-frame $U$ and observed $g$ (KiDS); rest-frame $V$ and observed $r$ (KiDS); and rest-frame $J$ and observed $H$ (VIKING). We note that for K-corrections to stay below a $\sim$5\% level, it is not
necessary to apply more detailed analyses of the underlying populations, as the method simply performs a sort of ``interpolation'' in the best-fit spectra to compute the colour transformations, and, given the redshifts covered, this interpolation is not stretched too much in wavelength. The UVJ colour diagram classifies nine objects as quiescent, namely the spheroids (3873542, 609701, 16143, 422365 and 693193) and the disks (288762, 486049, 723783 and 79071). However, a fair amount of scatter is also evident throughout the whole sample, with objects deeply in the star-forming region. Part of the (horizontal) observed scatter in this diagram may be due to the fact that the $H$-band imaging is shallower, with poorer resolution. 

\subsection{The role of environment}
\label{subsec:environment}

Environment is arguably one of the most important factors driving the 
evolution of compact massive galaxies with redshift. Within the framework
of the two-stage formation scenario \citep[e.g.][]{Oser:10}, massive
galaxies have an {\sl in-situ} formation channel, by which gas is typically
fed to the centre of the galaxy, triggering a star formation burst, and an
{\sl ex-situ} channel where stars are incorporated into the galaxy via mergers.
Minor mergers may be especially relevant in the evolution from massive cores
at high redshift to the present massive and large galaxies, as this type of
mergers tend to populate the outer regions of galaxies \citep{Naab09}.
This could lead to the formation of a spheroidal envelope \citep{Hopkins09b,Bezanson09} or the growth of a galaxy disk \citep{Graham15,delaRosa16}, but always resulting in an increased size and a subsequent evolution on the mass-size plane \citep{Buitrago17}. 

Therefore, one should expect the growth of massive
galaxies to be strongly dependent on environment, reflecting their past merger history.
\citet{Valentinuzzi10a,Valentinuzzi10b} found a rather high prevalence of massive compact
galaxies in low-z clusters, amounting to $\sim 20$\% of their total sample
of cluster galaxies with stellar mass above $3\times 10^{10}$M$_\odot$.
This result is in stark contrast with the 4.4\% equivalent fraction 
found in the field \citep{Poggianti13}.
Moreover, the clearest example of a local compact and massive relic
is NGC\,1277, situated in the Perseus cluster \citep{vandenBosch12,Ferre-Mateu17}.
By definition, low-redshift relics should represent 
galaxies that formed a very massive core early on (at redshift z$\gtrsim$2-3), 
and did not experience any significant subsequent merging -- or alternatively
underwent a dramatic dynamical event that removed its envelope at a later time.
Therefore, we can roughly separate two major channels for the formation of MUGs, 
based on environment, that will create two different populations: 
1) a high density sample representing galaxies that were incorporated into clusters
at early times, with two potential scenarios: either systems that did not undergo any merger, given the high relative velocities, or non-compact massive galaxies that were stripped of their outer envelope via tidal interactions in the cluster; and 
2) a low density sample, where these galaxies represent local and isolated high-density
peaks that allowed for the formation of a massive galaxy at early times, but where
the local density was not high enough to experience any mergers. The latter represents a cleaner, but, obviously, a much rarer sample than the former. Our GAMA-based selection
is the best option at present to target this question.

\begin{figure*}
\centering
    \resizebox{\hsize}{!}{\includegraphics{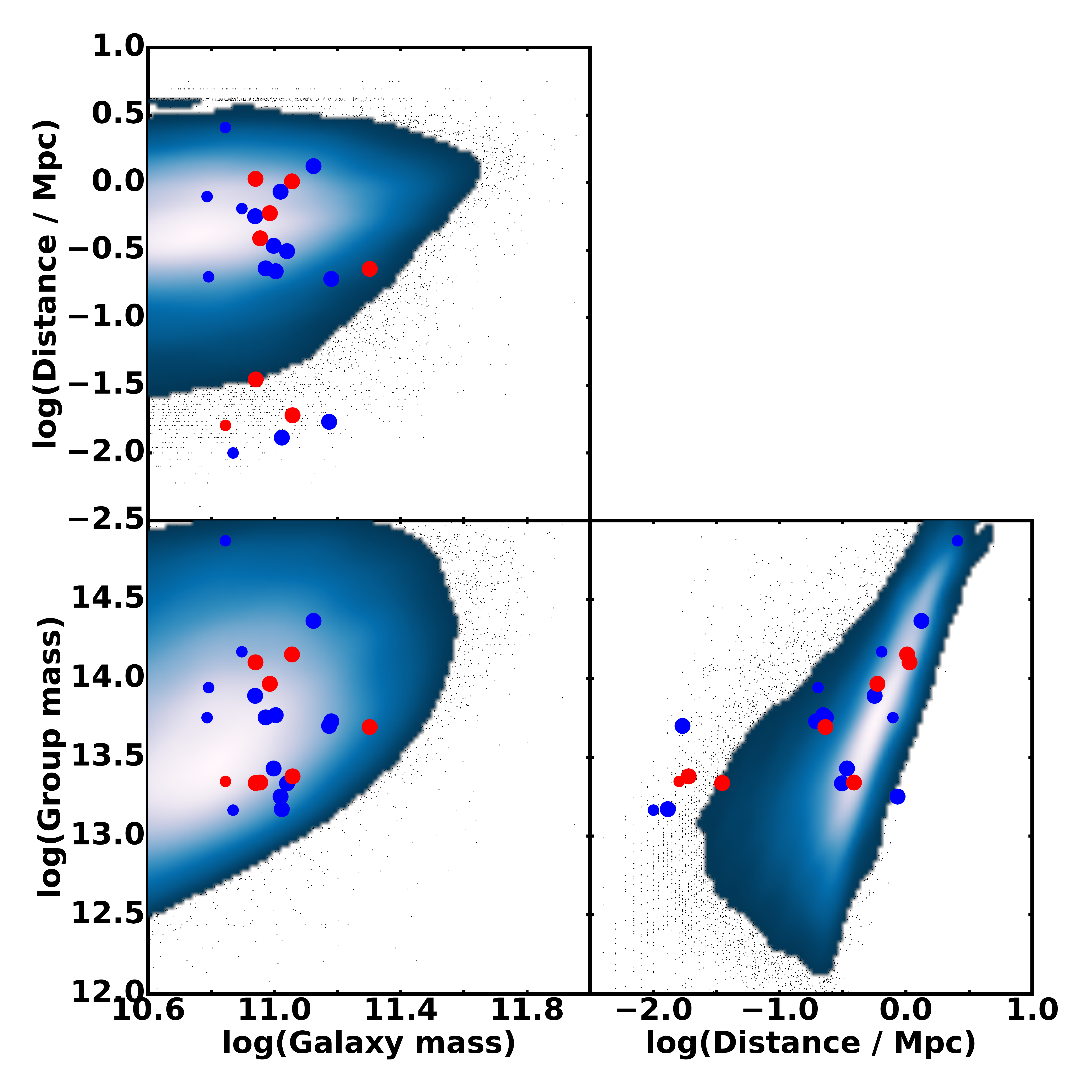}}
    \caption{The environment of our MUG sample is compared
      with the general distribution of GAMA massive galaxies. This is depicted by a blue colored Kernel Density Estimation, using individual datapoints where the density of datapoints is too low. Our MUG sample is
      colour and size coded as in Fig.~\ref{fig:Dn4K}, split 
      according to the S\'ersic index into disk-like (n$<$2.5, blue) and
      spheroid-like (n$>$2.5, red). The group dynamical masses and cluster-centric distances distances are
      taken from the G$^3$C catalogue of \citet{Robotham11}, see text for details.}
    \label{fig:enviro}
\end{figure*}

We contrast here our sample with the general distribution of massive galaxies
regarding group mass. The GAMA survey was especially designed
for the robust determination of galaxy environments, minimising as much as
possible the incompleteness of the spectroscopically-derived redshifts.
The group masses are taken from the G$^3$C catalogue of
\citet[][we use v10]{Robotham11}, following the scaling relation with
total luminosity from \citet{Viola15}. Fig.~\ref{fig:enviro} shows
the comparison, where our massive compact galaxies are shown as filled
circles -- following the same colour coding as in Fig~\ref{fig:Dn4K} -- while
the Kernel Density Estimation and the individual black dots 
represent the general population of massive galaxies in GAMA.
Only 17/6 of the 22/7 MUGs have a group allocation in the G$^3$C catalogue.
Most of our sample lies close to their group centre, but they do not reside in 
the most massive groups of the survey. Only four galaxies are located in groups more 
massive than $10^{14}$M$_\odot$, representing only 15\% of the total. 
The GAMA ``Environment Measures'' catalogue of \citet{Brough13} only lists 3 of
our MUG candidates\footnote{This catalogue is limited to z$_{\rm TONRY}<$0.18}, namely 79071 ($\Sigma_5$=2.58\,Mpc$^{-2}$);  
16143 ($\Sigma_5$=6.94\,Mpc$^{-2}$); and 609701 ($\Sigma_5$=8.53\,Mpc$^{-2}$),
where the numbers in brackets give the surface number densities estimated
within the 5th nearest neighbour. These kind of densities imply a distance to 
a 5th neighbour, with absolute magnitude brighter than M$_r<-20$, of <0.8Mpc.
We note that in the GAMA sample, the distribution
of $\Sigma_5$ has a mean of $2.68$\,Mpc$^{-2}$, and a median of $0.35$\,Mpc$^{-2}$.
Therefore, these three galaxies are clearly located in high density regions, although
not at the highest densities expected of galaxy clusters. 

Note that given the special character of this sample, and the relatively faint fluxes
probed with respect to the GAMA threshold (see~Tab.~\ref{tab:struct_param_r_band}), we may
wonder whether the completeness is high enough for these specific targets. For each detection,
GAMA provides a parameter ({\tt MASK\_IC}) that gives the completeness level within a region.
In our sample, the completeness levels are in most cases above 90\% except for two sources:
79071 (with 87\% completeness) and 319149 (88\%). How well are we able to detect the host environments of MUGs? A simple estimate based on simple stellar populations from the synthetic models of \citet{BC03} shows that at the highest redshift of our set 
(z$\sim$0.3), a solar-metallicity, dustless 10\,Gyr old population -- the oldest  possible at that redshift -- with stellar mass $10^{11}$M$_\odot$ would have an apparent magnitude r$\sim$19.8\,AB, i.e. the
limiting magnitude of the GAMA survey. Follow-up spectroscopy covering the fainter
sources surrounding the MUGs will be needed to understand the issue of environment in more detail.

\subsection{Number densities of ultracompact galaxies at 0.02$<$z$<$0.3}
\label{subsec:number_densities}

\begin{figure}
\centering
    \includegraphics[width=8cm]{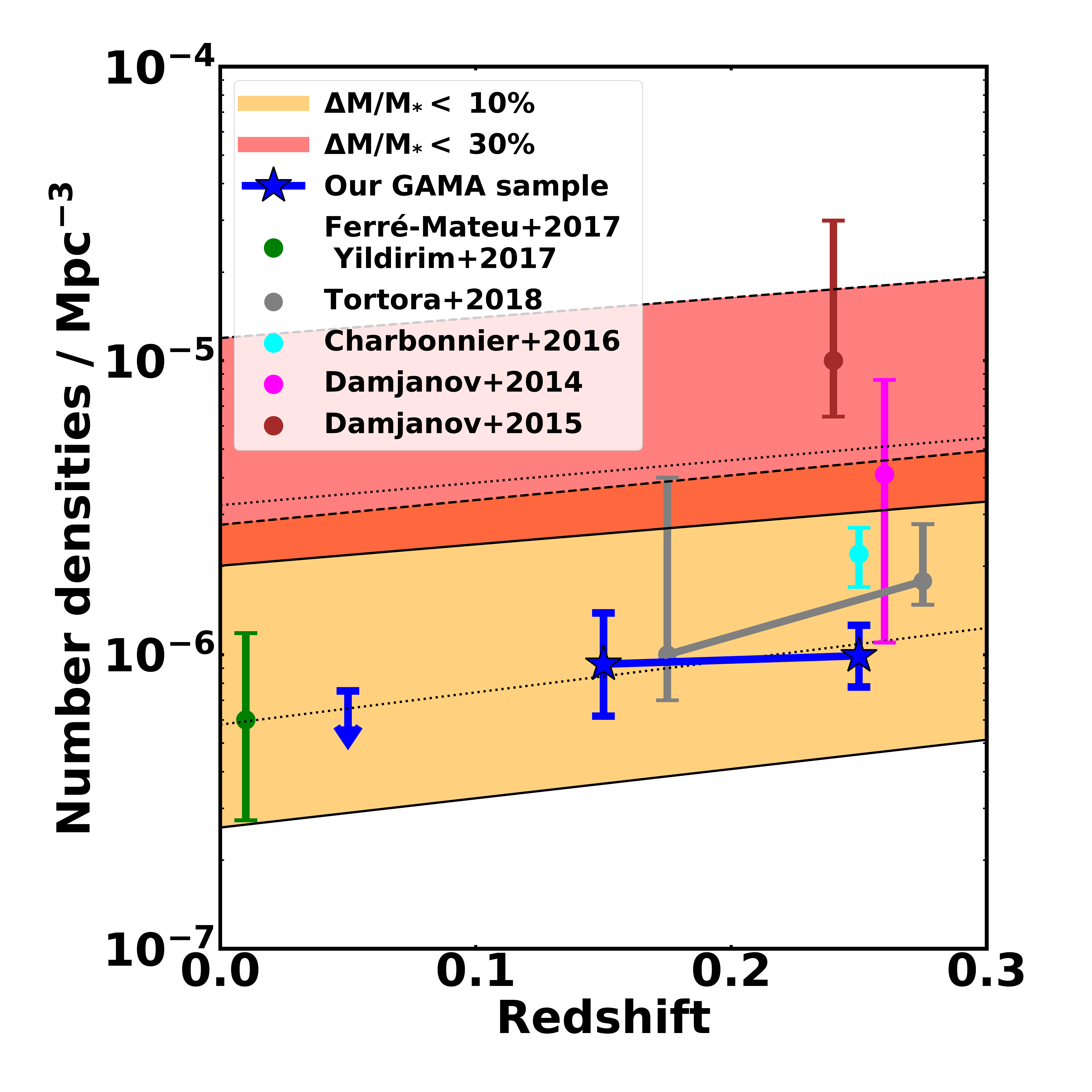}
    \caption{Redshift evolution of the MUG comoving number density over the redshift range  0.02$<$z$<$0.3, along with the theoretical expectations (shaded regions) taken from Fig. 2 in \citet{Quilis13}. See \ref{subsec:number_densities} for details. }
    \label{fig:number_densities}
\end{figure}

Fig.~\ref{fig:number_densities} shows the comoving number densities of MUGs at 0.02$<$z$<$0.3. Our data points are the blue stars, combining both our M$_{\star}$ $>$ 8$\times$10$^{10}$ M$_{\odot}$ and our  6$\times$10$^{10}$ $<$ M$_{\star}$/M$_{\odot}$ $<$ 8$\times$10$^{10}$ MUG subsamples. The inferred number density values are listed in Table \ref{tab:number_densities}. The shaded regions represent the results from \citet{Quilis13}, where they show the redshift evolution of the number density of massive ($>$ 8$\times$10$^{10}$M$_\odot$) galaxies whose stellar masses have been modified after their formation episode by less than 10\% (yellow area) and 30\% (red area), according to the simulations in \citet{deLucia07,Guo11,Guo13}.

\input{number_density_table.tex}

We did not find any MUG at z$<$0.1, but we can place an upper limit from our observations and also an additional data point at z$\sim$0 from the 3 relic galaxies in \citet{Ferre-Mateu17}. However, we note that, according to our definition of compactness, only NGC1277 and PGC32873 would be included in our sample (rejecting Mrk1216). We also contrast these results with those in \citet{Yildirim17}, where only 2 galaxies would pass our selection, NGC1277 and PGC70520 (discarding PGC32873 this time). Therefore, broadly speaking, there are 3 galaxies consistent with our definition of a MUG within a distance of 106\,Mpc, namely NGC1277, PGC32873 and PGC70520; and thus we keep the number density value of 6$\times$10$^{-7}$ Mpc$^{-3}$ at z$\sim$0 from \citet{Ferre-Mateu17}.

At z$>$0, there is a disparity in the data point values presented in our plot, because of the non homogeneity in the MUG definitions. We took in all cases the selection criteria that are closest to ours. At 0.2$<$z$<$0.3, we show:
\begin{itemize}
\item \citet{Charbonnier17}: M$_{\star}$ $>$ 10$^{10.7}$ M$_{\odot}$ and R$_{\rm e}$ $<$ 1.5 kpc·(M$_{\star}$/10$^{11}$ M$_{\odot}$)$^{0.75}$ ($i$-band imaging of SDSS/Stripe 82 with the CFHT)
\item \citet{Damjanov14}: M$_{\rm dyn}$ $>$ 8$\times$10$^{10}$ M$_{\odot}$ (stellar-like objects in SDSS data, classified as quiescent galaxies, for which dynamical masses are obtained).
\item \citet{Damjanov15}: M$_{\star}$ $>$ 8$\times$10$^{10}$ M$_{\odot}$ and R$_{\rm e}$ $<$ 2.5 kpc·(M$_{\star}$/10$^{11}$ M$_{\odot}$)$^{0.75}$ (ACS/COSMOS data of passive galaxies).
\item \citet{Tortora18}, superseding \citet{Tortora16}: M$_{\star}$ $>$ 8$\times$10$^{10}$ M$_{\odot}$ and R$_{\rm e,circ}$ $<$ 1.5 kpc (median between g-, r- and i-bands)
\end{itemize}

Summarizing, the disparity among different works comes from a combination of both slightly different selection criteria and the scarcity of this type of galaxy. Given the uncertainties from the different studies, we could affirm that the number density of these objects is $\sim$10$^{-6}$ Mpc$^{-3}$ at 0$<$z$<$0.3. The error bars in our data are consistent with either a flat slope or a weak redshift evolution, as suggested by the coloured areas in Fig. 14.


\section{Conclusions}
\label{sec:conclusions}

We present in this paper a complete census within the footprint of the GAMA survey of Massive Ultracompact Galaxies (MUGs) at 0.02$<$z$<$0.3, placing special emphasis on their number density and environmental determination. GAMA is the perfect benchmark for this study due to its spatially uniform spectroscopic completeness ($\sim$98.5\% down to $r =$ 19.8 ABmag) over a large area of the sky (180 deg$^{2}$) with deep ancillary photometric data ($g$-, $r$- and $i$-bands from KiDS; $Z$-band from VIKING). This imaging enabled us to produce robust size estimates, as it is $\sim$2\,mag deeper than SDSS and with better spatial resolution ($<$0.7\,arcsec seeing in the $r$-band).

Our final sample consists of 22 objects with effective radii $<$ 2\,kpc in at least two photometric bands and stellar masses M$_{\star}>8\times10^{10}$M$_{\odot}$. We include an additional set of 7 MUGs if taking into account our mass uncertainties, and a potentially heavier initial mass function within this type of galaxies. This work builds up from previous low-redshift size determinations in SDSS data, conducting a careful analysis of pre-selected compact galaxies, using deeper imaging and simulations to assess the reliability of our results.

MUGs are located, by definition, at the bottom of the galaxy mass-size relation (see Fig.~\ref{fig:mass_size}), but interestingly they usually display early disk/swollen disk morphologies (Fig.~\ref{fig:sersic_r}) and very large stellar densities (Fig.~\ref{fig:hist_densities_in_2D}). Some of them host very old and passive stellar populations (Figures \ref{fig:Dn4K} and \ref{fig:uvj}), therefore being ``red nugget'' galaxies that survive unaffected over a large fraction of the present age of the Universe. Hence, the study of this population (at low redshift) is a true window to the early phases of galaxy evolution.

State-of-the-art simulations suggest that relic galaxies should inhabit the central regions of galaxy clusters in order not to have experienced galaxy mergers -- due to their inherent large relative velocities -- and thus maintain their compact sizes across cosmic time. A confirmation of the above scenario is only possible by means of a high-completeness, large-area spectroscopic survey, such as GAMA. However, given the extremely low number density of these objects, a wider coverage -- at similar levels of completeness -- would be required to unambiguously determine the channels that lead to massive compact galaxies at low redshift. Nevertheless, we can state beyond doubt (Section \ref{subsec:environment}, Figure \ref{fig:enviro}) that a sizeable number of our massive compact galaxies do not reside in overdense environments, at odds with theoretical predictions.

Our sample shows no trend with group mass. Looking at dark matter halo mergers only, \citet{Fakhouri10} have shown that the average number of mergers a halo encounters is both a strong function of the amount of elapsed time and the final mass of the halo. More massive groups should have undergone more mergers on average. In light of this, our results show an opposite trend if we identify MUGs as proxies for a lack of merger activity in the past. As MUGs are almost equally distributed among groups of different masses (again see Section \ref{subsec:environment}, Figure \ref{fig:enviro}),  the relative fraction of MUGs is higher in massive groups, since they are exponentially less abundant than low mass groups. This implies two possible scenarios, (1) either the sample of MUGs reside in host halos that underwent an atypical growth history, or (2) dynamical effects associated with galaxy mergers play an important role. It is to be expected that both of these will play a role. Theoretical models following the merger of galaxies within merging halos using dynamical friction estimates have shown that the the size evolution of massive early-type galaxies with M$_{\star}> 5 \times$10$^{10}$ M$_{\odot}$ are to first order independent of the halo/group mass for massive galaxies \citep[e.g.][]{Khochfar06}, which would support that part of the observed trend could be driven by the merger process of galaxies once their halos have merged. It is well established that once the velocity dispersion of galaxies in massive clusters becomes too high, merging will effectively cease to happen.

The above considerations lead to the following scenarios for the origin of MUGs: a) the passive sub-sample of early-types most-likely formed in halos that showed a fast growth history at high redshift and assembled enough mass quickly to increase the velocity dispersion of galaxies within the halo hindering mergers, especially minor mergers; b) the population of compact disc-like systems with ongoing star formation, in contrast, resides in halos that assembled at later times and, thus, merging of galaxies did not have enough time to proceed, or the environment may prevent mergers, depending on the group mass.

Regarding the comoving number densities of MUGs, the combination of our results with previous work from the literature suggest that, 
at 0$<$z$<$0.3, these objects have number densities $\sim$10$^{-6}$\,Mpc$^{-3}$ (see Tab.~3 and Fig. \ref{fig:number_densities}). This value indicates that this galaxy population is very rare, and also imposes an upper limit on the density of relic galaxies, which, by definition, (massive and small vs massive and small and old) must be even less numerous. Remarkably, the MUG/``red nugget'' number density is $\sim$10$^{-4}$ Mpc$^{-3}$ at $z =$2-3, implying that less than one in a hundred of them survive untouched from their formation epoch. When comparing these results with simulations on a density vs redshift
diagram (see Fig.~\ref{fig:number_densities}), the data points lay in a region where the galaxies increased their stellar mass by less than 10\% across cosmic time. This result illustrates the uniqueness of these galaxies, suggesting a minimal contribution from merging events.
Upcoming surveys such as WAVES \citep{WAVES}, covering wider regions of the sky, and deeper in flux than GAMA, will
allow us to produce a high enough number of MUGs, to explore their formation channels in detail.

It is remarkable that our sample of massive compact galaxies appear -- at least in projection -- in a relatively crowded environment (see Figs~\ref{fig:mosaic_rgb} and \ref{fig:mosaic_rgb_borderline}), complicating the photometric and spectroscopic analysis. Therefore, the inferred sizes and densities only provide upper bounds, as the light from the minor companions may contaminate the surface brightness profiles in a way that is impossible to mitigate at the spatial 
resolutions available to us. Follow-up observations, including imaging at high spatial resolution with HST or JWST, and deeper spectroscopy with IFU cameras at 8-10m-class telescopes, such as MUSE or MEGARA,  will be required to better comprehend the intriguing properties of massive ultracompact galaxies.

\begin{acknowledgements}
FB acknowledges the support by FCT via the postdoctoral fellowship SFRH/BPD/103958/2014. This work is supported by Funda\c{c}\~ao para a Ci\^encia e a Tecnologia (FCT) through national funds (UID/FIS/04434/2013) and by FEDER through COMPETE2020 (POCI-01-0145-FEDER-007672). FB also acknowledges support from grant AYA2016-77237-C3-1-P from the Spanish Ministry of Economy and Competitiveness (MINECO). K. Kuijken and M. Viola kindly provided us with the shapelet PSFs for our photometric analysis.
I. Trujillo and V. Quilis kindly provided us with the data of fig.~2 from \citet{Quilis13}. We also acknowledge the help of A. Molino, I. Matute, S. Reis and P. Norberg. This research made use of APLpy, an open-source plotting package for Python \citep{aplpy_ref}.
We have used extensively the following software packages: TOPCAT \citep{Taylor05}, ALADIN \citep{Bonnarel2000} and MATPLOTLIB \citep{matplotlib_ref}.
GAMA is a joint European-Australasian project based around a spectroscopic
campaign using the AAT. The GAMA input catalogue is based on data
taken from the SDSS and the UKIRT Infrared Deep Sky
Survey. Complementary imaging of the GAMA regions is being obtained by
a number of independent survey programmes including GALEX MIS, VST
KIDS, VISTA VIKING, WISE, Herschel-ATLAS, GMRT and ASKAP providing UV
to radio coverage. GAMA is funded by the STFC (UK), the ARC
(Australia), the AAO and the participating institutions. The GAMA web
site is http://www.gama-survey.org/.
\end{acknowledgements}


\bibliographystyle{aa.bst}
\bibliography{refs.bib}


\begin{appendix}

\section{Robustness of the measured structural parameters}
\label{app:simus_robustness_struct_params}

We simulated 3840 galaxies, both in the KiDS $r$-band and the VIKING $Z$-band, with the tools described in \citet{Buitrago13}. We explored the range of structural parameters shown by our sample, namely:

$$\rm 15 < mag < 20$$
$$\rm 1 < R_{\rm e}/pix < 10$$
$$1 < n < 8$$
$$\rm 0.3 < ar < 1$$
$$\rm 0 < pa < 90$$

where mag, R$_{\rm e}$, $n$, ar and pa stand out for magnitude (either in the $r$- or $Z$-band), effective radius, S\'ersic index, axis ratio and position angle.

The structural parameters of the mock galaxies were randomly distributed in a linear way, considering the full parameter space defined by the previously mentioned values. The mock galaxies were placed randomly on the KiDS and VIKING images, only imposing the condition that they should not overlap with any detection defined by the SExtractor segmentation map of each survey. Each mock galaxy was convolved with a representative PSF from those images, and analyzed with the same code we utilize to investigate the real galaxies.

\begin{figure*}
\centering
    \includegraphics[width=17cm]{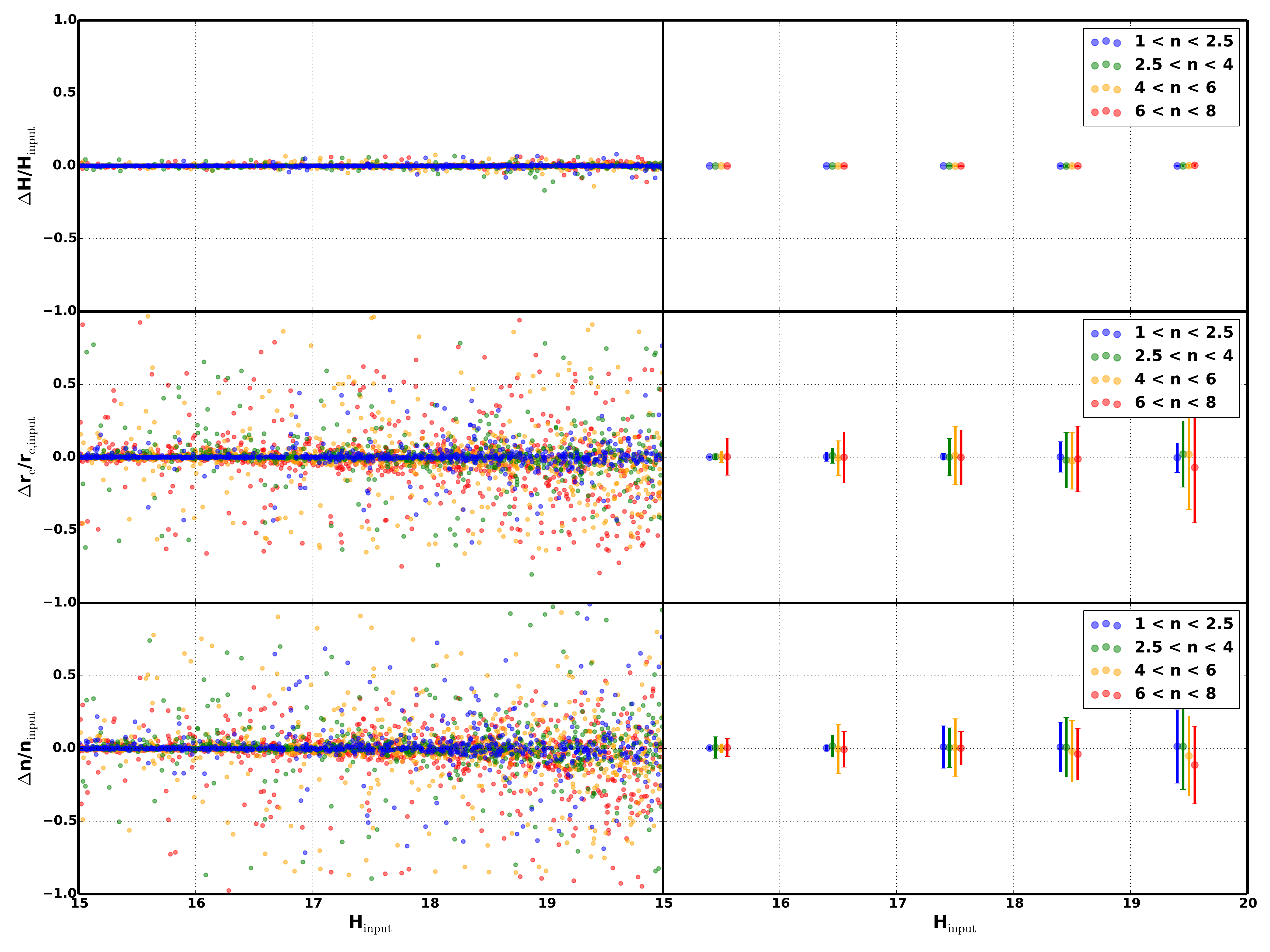}
    \caption{Relative errors – (output-input)/input – of the structural parameters (magnitude, effective radius and S\'ersic index) of our simulated KiDS galaxies in the $r$-band. The right-hand column shows the average values in 1\,mag bins (derived after a 5$\sigma$ clipping). The error bars represent the standard deviation within each bin.}
    \label{fig:sims_r_band_mag}
\end{figure*}

\begin{figure*}
\centering
    \includegraphics[width=17cm]{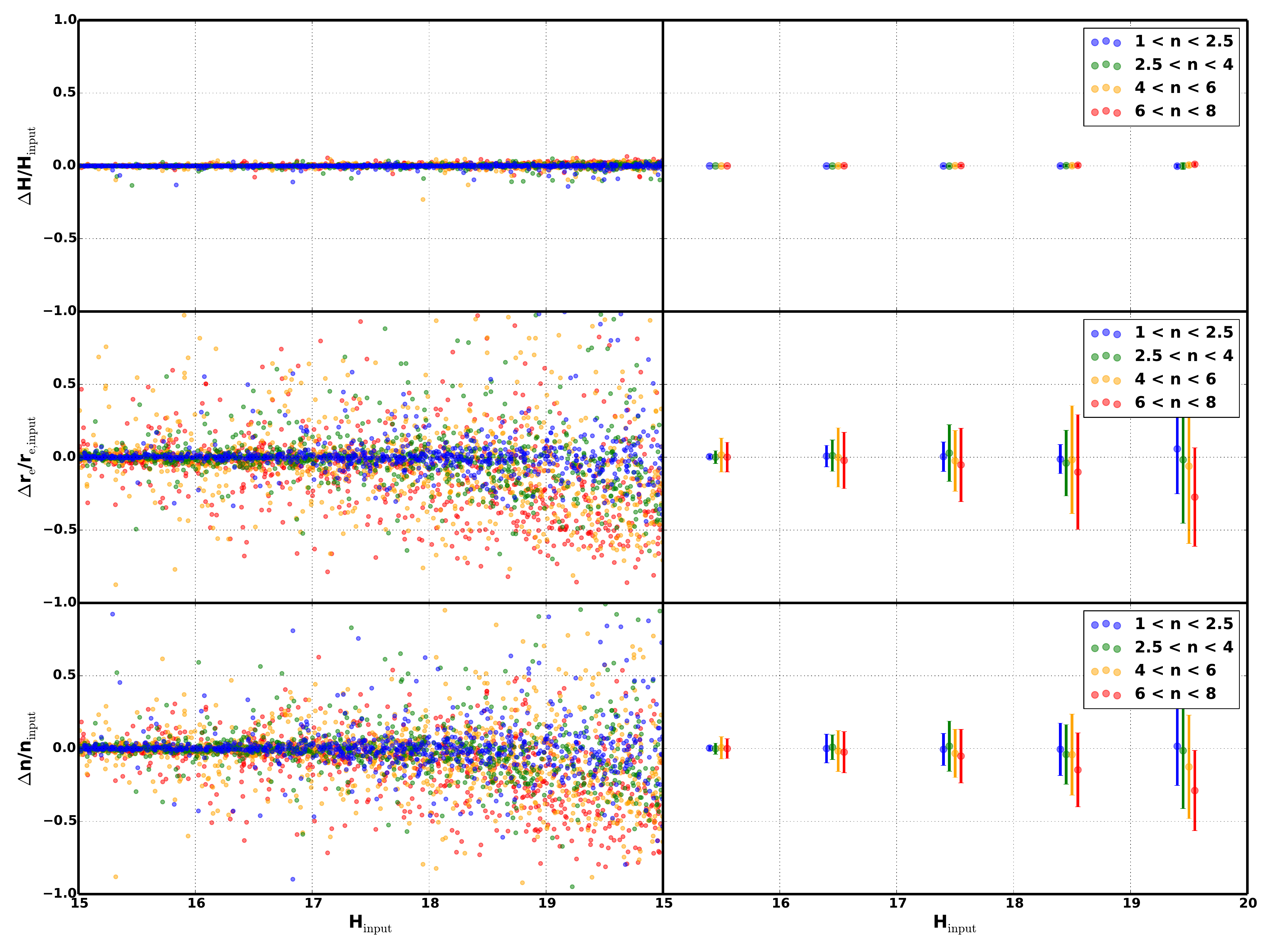}
    \caption{Relative errors – (output-input)/input – of the structural parameters (magnitude, effective radius and S\'ersic index) of our simulated VIKING galaxies in the $Z$-band. The right-hand column shows the average values in 1\,mag bins (derived after a 5$\sigma$ clipping). The error bars represent the standard deviation within each bin.}
    \label{fig:sims_z_band_mag}
\end{figure*}

According to Figures~\ref{fig:sims_r_band_mag} and \ref{fig:sims_z_band_mag}, our ability to recuperate the structural parameters degrades at fainter fluxes and higher S\'ersic indices. This result is expected, i.e. dimmer and more concentrated objects are more difficult to analyze \citep[a similar behaviour is found in the simulations of][]{Buitrago13}. In the $Z$-band, the results are somewhat worse because of the coarser pixel scale (0.21\,arcsec/pix versus 0.339\,arcsec/pix). However, for the data at hand, caution needs to be taken only for faint objects with a high S\'ersic index.

\section{Structural parameters for the g-, i- and Z-bands}
\label{app:struct_params_g_r_i}

\input{struct_param_table_g_band.tex}
~\\

\input{struct_param_table_i_band.tex}
~\\

\input{struct_param_table_z_band.tex}
~\\

\end{appendix}

\end{document}

%% file: table_main_props.tex
\begin{table*}
\centering
\label{tab:main_props}
\caption{Basic observational properties of our Massive Ultracompact Galaxy (MUG) sample from GAMA.}
\begin{tabular}{cccccccc}
\hline \hline
CATAID & RA & DEC & z$_{\rm spec}$ & log$_{10}$ M$_{\rm stellar}$ & log$_{10}$ $\langle\Sigma_{\rm 1.5}\rangle$ & log$_{10}$ $\langle\Sigma_{e}\rangle$ & log$_{10}$ $\langle\rho_{e}\rangle$ \\
\hline
79071 & 219.96358837 & 0.11254047 & 0.1335 & 11.00 & 10.57 & 9.63 & 9.63 \\
16143 & 217.73607469 & 0.74513527 & 0.1378 & 10.99 & 10.57 & 9.63 & 9.23 \\
784327 & 178.74986841 & -1.42614761 & 0.1625 & 11.12 & 11.00 & 10.16 & 10.16 \\
609701 & 176.87118407 & 0.31908043 & 0.1774 & 10.94 & 10.54 & 9.61 & 9.22 \\
493286 & 220.3298223 & -1.28795637 & 0.1849 & 11.00 & 10.94 & 10.12 & 10.12 \\
319149 & 213.61987948 & 1.85908419 & 0.1881 & 11.30 & 10.94 & 10.03 & 9.66 \\
84466 & 178.0554695 & 0.56305098 & 0.1973 & 11.18 & 10.91 & 10.02 & 10.02 \\
178241 & 179.75149717 & -1.93248107 & 0.2059 & 11.02 & 10.51 & 9.54 & 9.54 \\
93202 & 219.52825636 & 0.53898409 & 0.2104 & 10.94 & 10.61 & 9.70 & 9.36 \\
63726 & 216.76230139 & -0.29771 & 0.2223 & 11.06 & 10.77 & 9.87 & 9.55 \\
765033 & 223.44512218 & 0.14390729 & 0.2360 & 10.95 & 10.61 & 9.69 & 9.33 \\
4220443 & 181.69997991 & -1.98105929 & 0.2541 & 11.11 & 10.99 & 10.14 & 9.94 \\
422365 & 130.57463302 & 2.59286311 & 0.2582 & 11.53 & 11.05 & 10.09 & 9.64 \\
71471 & 184.92124534 & 0.02897459 & 0.2592 & 11.04 & 10.89 & 10.05 & 10.05 \\
3873542 & 129.17293417 & -1.27914494 & 0.2609 & 11.19 & 10.81 & 9.89 & 9.51 \\
300986 & 131.463188 & 1.15208397 & 0.2663 & 11.02 & 11.29 & 10.58 & 10.58 \\
221269 & 185.47490703 & 1.6248479 & 0.2671 & 10.94 & 10.53 & 9.60 & 9.60 \\
288762 & 180.01770455 & 1.78508059 & 0.2751 & 11.17 & 10.81 & 9.89 & 9.89 \\
138954 & 184.98809975 & -1.74957511 & 0.2767 & 11.14 & -- & -- & -- \\
855304 & 131.15298472 & 1.36284576 & 0.2879 & 10.97 & 11.18 & 10.45 & 10.45 \\
791716 & 181.36721636 & -1.46390455 & 0.2886 & 11.06 & -- & -- & -- \\
693193 & 223.21201479 & -0.3199314 & 0.2989 & 11.16 & 10.87 & 9.98 & 9.66 \\
\hline \hline
723783 & 223.09000412 & 2.4662566 & 0.1975 & 10.83 & 11.07 & 10.35 & 10.35 \\
486049 & 218.39660186 & -1.74897733 & 0.2242 & 10.90 & 10.46 & 9.51 & 9.51 \\
388238 & 139.19204138 & 2.54304803 & 0.2291 & 10.79 & 10.66 & 9.82 & 9.82 \\
55006 & 180.21098968 & -0.39229633 & 0.2603 & 10.84 & 10.52 & 9.61 & 9.27 \\
750374 & 214.89578661 & 0.6837029 & 0.2799 & 10.87 & 10.68 & 9.82 & 9.82 \\
3173601 & 182.42594704 & 1.27935477 & 0.2874 & 10.79 & 10.80 & 10.01 & 10.01 \\
365375 & 140.28918164 & 2.62257045 & 0.2874 & 10.85 & 10.49 & 9.57 & 9.57 \\
\hline
\end{tabular}
\tablefoot{The division line splits the galaxies with stellar mass greater or lower than 8$\times$10$^{10}$M$_{\odot}$. Columns: (1) GAMA ID, (2) Right Ascension (J2000), (3) Declination (J2000), (4) Spectroscopic redshift, (5) Stellar mass, in units of log(M$_{\odot}$), (6) 2D stellar surface density according to the prescriptions in \citet{Barro13}, in units of log(M$_{\odot}$ kpc$^{-1.5}$), (7) 2D stellar surface density in units of log(M$_{\odot}$ kpc$^{-2}$), (8) 3D stellar density in units of log(M$_{\odot}$ kpc$^{-3}$). The quantities in cols. (6-8) are averaged within the effective radius.}
\end{table*}

%% file: struct_param_table_r_band.tex
\begin{table*}
\centering
\label{tab:struct_param_r_band}
\caption{List with the $r$-band structural parameters of our MUG sample.}
\begin{tabular}{cccccc}
\hline \hline
CATAID & mag $\pm \delta$mag & r$_{\rm e} \pm \delta$r$_{\rm e}$ & r$_{\rm e,circ} \pm \delta$r$_{\rm e,circ}$ & n $\pm~\delta$n & ar $\pm~\delta$ar \\
 & & [kpc] & [kpc] & & \\
\hline
79071 & 16.93 $\pm$  0.01 & 1.92 $\pm$ 0.01 & 1.67 $\pm$ 0.11 & 2.08 $\pm$ 0.05 & 0.76 $\pm$ 0.10 \\
16143 & 18.44 $\pm$  0.01 & 1.89 $\pm$ 0.06 & 1.85 $\pm$ 0.20 & 5.68 $\pm$ 0.19 & 0.96 $\pm$ 0.15 \\
784327 & 19.38 $\pm$  0.12 & 1.21 $\pm$ 0.22 & 1.08 $\pm$ 0.21 & 2.19 $\pm$ 0.44 & 0.80 $\pm$ 0.03 \\
609701 & 19.51 $\pm$  0.09 & 1.85 $\pm$ 0.46 & 1.67 $\pm$ 0.57 & 3.75 $\pm$ 0.75 & 0.81 $\pm$ 0.14 \\
493286 & 19.99 $\pm$  0.09 & 1.11 $\pm$ 0.07 & 1.04 $\pm$ 0.25 & 1.73 $\pm$ 0.62 & 0.88 $\pm$ 0.30 \\
319149 & 19.34 $\pm$  0.05 & 1.73 $\pm$ 0.22 & 1.33 $\pm$ 0.35 & 4.12 $\pm$ 0.46 & 0.59 $\pm$ 0.16 \\
84466 & 19.22 $\pm$  0.19 & 1.52 $\pm$ 0.24 & 1.45 $\pm$ 0.39 & 1.62 $\pm$ 0.82 & 0.91 $\pm$ 0.20 \\
178241 & 19.26 $\pm$  0.40 & 2.21 $\pm$ 0.46 & 2.08 $\pm$ 0.48 & 1.64 $\pm$ 1.13 & 0.89 $\pm$ 0.03 \\
93202 & 17.67 $\pm$  0.01 & 1.66 $\pm$ 0.01 & 1.42 $\pm$ 0.09 & 5.11 $\pm$ 0.09 & 0.74 $\pm$ 0.10 \\
63726 & 19.91 $\pm$  0.05 & 1.57 $\pm$ 0.19 & 1.39 $\pm$ 0.46 & 4.03 $\pm$ 0.45 & 0.79 $\pm$ 0.33 \\
765033 & 19.86 $\pm$  0.01 & 1.71 $\pm$ 0.01 & 0.80 $\pm$ 0.17 & 6.07 $\pm$ 0.13 & 0.22 $\pm$ 0.09 \\
4220443 & 19.96 $\pm$  0.01 & 1.21 $\pm$ 0.27 & 0.67 $\pm$ 0.30 & 4.66 $\pm$ 1.10 & 0.31 $\pm$ 0.13 \\
422365 & 19.94 $\pm$  0.10 & 2.11 $\pm$ 0.53 & 1.99 $\pm$ 0.66 & 3.15 $\pm$ 0.63 & 0.89 $\pm$ 0.15 \\
71471 & 19.34 $\pm$  0.09 & 1.25 $\pm$ 0.08 & 1.16 $\pm$ 0.14 & 0.92 $\pm$ 0.33 & 0.86 $\pm$ 0.09 \\
3873542 & 18.63 $\pm$  0.02 & 1.80 $\pm$ 0.06 & 1.21 $\pm$ 0.36 & 3.21 $\pm$ 0.08 & 0.45 $\pm$ 0.24 \\
300986 & 19.43 $\pm$  0.09 & 0.66 $\pm$ 0.04 & 0.32 $\pm$ 0.11 & 1.65 $\pm$ 0.59 & 0.23 $\pm$ 0.13 \\
221269 & 19.63 $\pm$  0.20 & 1.86 $\pm$ 0.30 & 1.37 $\pm$ 0.56 & 1.16 $\pm$ 0.59 & 0.54 $\pm$ 0.27 \\
288762 & 20.01 $\pm$  0.09 & 1.75 $\pm$ 0.11 & 1.11 $\pm$ 0.26 & 1.61 $\pm$ 0.58 & 0.40 $\pm$ 0.14 \\
138954 & 19.28 $\pm$  0.03 & - & - & - & - \\
855304 & 20.93 $\pm$  0.09 & 0.73 $\pm$ 0.05 & 0.35 $\pm$ 0.07 & 1.40 $\pm$ 0.50 & 0.23 $\pm$ 0.07 \\
791716 & 20.10 $\pm$  0.01 & - & - & - & - \\
693193 & 18.22 $\pm$  0.01 & 1.55 $\pm$ 0.03 & 1.43 $\pm$ 0.31 & 5.50 $\pm$ 0.05 & 0.85 $\pm$ 0.34 \\
\hline \hline
723783 & 18.21 $\pm$  0.08 & 0.69 $\pm$ 0.13 & 0.44 $\pm$ 0.20 & 0.83 $\pm$ 0.11 & 0.41 $\pm$ 0.21 \\
486049 & 19.70 $\pm$  0.20 & 1.96 $\pm$ 0.31 & 1.83 $\pm$ 0.52 & 1.40 $\pm$ 0.71 & 0.87 $\pm$ 0.22 \\
388238 & 18.66 $\pm$  0.09 & 1.22 $\pm$ 0.23 & 1.14 $\pm$ 0.37 & 1.82 $\pm$ 0.23 & 0.87 $\pm$ 0.24 \\
55006 & 19.43 $\pm$  0.01 & 1.65 $\pm$ 0.37 & 1.43 $\pm$ 0.59 & 4.34 $\pm$ 1.02 & 0.75 $\pm$ 0.29 \\
750374 & 19.75 $\pm$  0.09 & 1.34 $\pm$ 0.09 & 1.24 $\pm$ 0.20 & 1.30 $\pm$ 0.46 & 0.86 $\pm$ 0.17 \\
3173601 & 19.54 $\pm$  0.09 & 0.98 $\pm$ 0.06 & 0.72 $\pm$ 0.13 & 1.98 $\pm$ 0.71 & 0.54 $\pm$ 0.12 \\
365375 & 20.05 $\pm$  0.09 & 1.74 $\pm$ 0.11 & 1.68 $\pm$ 0.01 & 0.92 $\pm$ 0.33 & 0.93 $\pm$ 0.01 \\
\hline
\end{tabular}
\tablefoot{The division line splits the galaxies with stellar mass greater or lower than 8$\times$10$^{10}$M$_{\odot}$. Columns: (1) GAMA ID, (2) Magnitude (3) Effective radius in kpc (4) Circularized effective radius in kpc (5) S\'ersic index (6) Axis ratio.}
\end{table*}

%% file: number_density_table.tex
\begin{table*}
\centering
\label{tab:number_densities}
\caption{Comoving number density of MUGs. M$_{10}$ is defined as the stellar mass of the
galaxy in units of $10^{10}$M$_\odot$. Error bars stem from the calculations in \citet{Gehrels1986} for low number statistics}
\begin{tabular}{ccccccc}
\hline \hline
 & All galaxies & Num. density & M$_{10}>$8 & Num. density & 6$<$M$_{10}<$8 & Num. density \\
  & \#Objects & $\times$10$^{-7}$ Mpc$^{-3}$ & \#Objects & $\times$10$^{-7}$ Mpc$^{-3}$ & \#Objects & $\times$10$^{-7}$ Mpc$^{-3}$ \\
\hline
0.02 $<$z$<$ 0.1 & 0 & $<$ 7.53 & - & - & - & - \\
0.1 $<$z$<$ 0.2 & 8  & 9.29$^{+4.58}_{-3.11}$ & 7  & 8.12$^{+4.38}_{-3.00}$ & 1 & 1.16$^{+2.67}_{-0.96}$ \\
0.2 $<$z$<$ 0.3 & 21 & 9.92$^{+2.67}_{-2.15}$ & 15 & 7.08$^{+2.34}_{-1.81}$ & 6 & 2.83$^{+1.69}_{-1.12}$ \\
\hline
\end{tabular}
\end{table*}

%% file: struct_param_table_g_band.tex
\begin{table*}
\centering
\label{tab:struct_param_g_band}
\caption{List with the $g$-band structural parameters for our MUG sample}
\begin{tabular}{cccccc}
\hline \hline
CATAID & mag $\pm \delta$mag & r$_{\rm e} \pm \delta$r$_{\rm e}$ & r$_{\rm e,circ} \pm \delta$r$_{\rm e,circ}$ & n $\pm~\delta$n & ar $\pm~\delta$ar \\
 & & [kpc] & [kpc] & & \\
\hline
79071 & 17.83 $\pm$  0.01 & 2.37 $\pm$ 0.01 & 1.98 $\pm$ 0.14 & 2.13 $\pm$ 0.05 & 0.70 $\pm$ 0.10 \\
16143 & 19.65 $\pm$  0.01 & 1.54 $\pm$ 0.06 & 1.34 $\pm$ 0.18 & 3.87 $\pm$ 0.19 & 0.76 $\pm$ 0.15 \\
784327 & 20.51 $\pm$  0.12 & 1.21 $\pm$ 0.22 & 1.11 $\pm$ 0.22 & 2.21 $\pm$ 0.44 & 0.85 $\pm$ 0.03 \\
609701 & 20.79 $\pm$  0.09 & 1.54 $\pm$ 0.46 & 1.51 $\pm$ 0.57 & 1.62 $\pm$ 0.75 & 0.97 $\pm$ 0.14 \\
493286 & 21.22 $\pm$  0.09 & 1.34 $\pm$ 0.07 & 1.26 $\pm$ 0.29 & 1.75 $\pm$ 0.62 & 0.87 $\pm$ 0.30 \\
319149 & 20.38 $\pm$  0.05 & 2.57 $\pm$ 0.22 & 2.20 $\pm$ 0.43 & 2.62 $\pm$ 0.46 & 0.73 $\pm$ 0.16 \\
84466 & 20.69 $\pm$  0.19 & 1.43 $\pm$ 0.24 & 1.22 $\pm$ 0.37 & 1.42 $\pm$ 0.82 & 0.72 $\pm$ 0.20 \\
178241 & 20.25 $\pm$  0.40 & 2.58 $\pm$ 0.46 & 2.44 $\pm$ 0.49 & 3.03 $\pm$ 1.13 & 0.90 $\pm$ 0.03 \\
93202 & 18.60 $\pm$  0.01 & 2.27 $\pm$ 0.01 & 2.12 $\pm$ 0.12 & 1.80 $\pm$ 0.09 & 0.88 $\pm$ 0.10 \\
63726 & 21.10 $\pm$  0.05 & 3.24 $\pm$ 0.19 & 2.48 $\pm$ 0.85 & 7.01 $\pm$ 0.45 & 0.59 $\pm$ 0.33 \\
765033 & 20.92 $\pm$  0.01 & 3.09 $\pm$ 0.01 & 2.05 $\pm$ 0.22 & 0.32 $\pm$ 0.13 & 0.44 $\pm$ 0.09 \\
4220443 & 21.34 $\pm$  0.01 & 1.64 $\pm$ 0.27 & 1.21 $\pm$ 0.35 & 2.31 $\pm$ 1.10 & 0.55 $\pm$ 0.13 \\
422365 & 21.36 $\pm$  0.10 & 1.72 $\pm$ 0.53 & 1.72 $\pm$ 0.65 & 3.41 $\pm$ 0.63 & 1.00 $\pm$ 0.15 \\
71471 & 20.59 $\pm$  0.09 & 1.70 $\pm$ 0.08 & 1.42 $\pm$ 0.16 & 5.44 $\pm$ 0.33 & 0.71 $\pm$ 0.09 \\
3873542 & 20.20 $\pm$  0.02 & 0.87 $\pm$ 0.06 & 0.82 $\pm$ 0.16 & 2.77 $\pm$ 0.08 & 0.89 $\pm$ 0.24 \\
300986 & 20.83 $\pm$  0.09 & 0.69 $\pm$ 0.04 & 0.57 $\pm$ 0.09 & 2.39 $\pm$ 0.59 & 0.67 $\pm$ 0.13 \\
221269 & 20.52 $\pm$  0.20 & 1.76 $\pm$ 0.30 & 1.33 $\pm$ 0.53 & 9.69 $\pm$ 0.59 & 0.57 $\pm$ 0.27 \\
288762 & 21.42 $\pm$  0.09 & 2.71 $\pm$ 0.11 & 2.58 $\pm$ 0.30 & 1.13 $\pm$ 0.58 & 0.90 $\pm$ 0.14 \\
138954 & 20.64 $\pm$  0.03 & - & - & - & - \\
855304 & 22.23 $\pm$  0.09 & 0.94 $\pm$ 0.05 & 0.61 $\pm$ 0.08 & 6.51 $\pm$ 0.50 & 0.41 $\pm$ 0.07 \\
791716 & 21.54 $\pm$  0.01 & 0.99 $\pm$ 0.01 & 0.32 $\pm$ 0.36 & 1.58 $\pm$ 0.01 & 0.11 $\pm$ 0.23 \\
693193 & 19.32 $\pm$  0.01 & 3.67 $\pm$ 0.03 & 2.76 $\pm$ 0.85 & 4.17 $\pm$ 0.05 & 0.57 $\pm$ 0.34 \\
\hline \hline
723783 & 19.56 $\pm$  0.08 & 1.76 $\pm$ 0.13 & 1.49 $\pm$ 0.33 & 1.16 $\pm$ 0.11 & 0.72 $\pm$ 0.21 \\
486049 & 21.02 $\pm$  0.20 & 2.51 $\pm$ 0.31 & 2.11 $\pm$ 0.58 & 1.58 $\pm$ 0.71 & 0.71 $\pm$ 0.22 \\
388238 & 19.96 $\pm$  0.09 & 1.30 $\pm$ 0.23 & 1.08 $\pm$ 0.38 & 2.12 $\pm$ 0.23 & 0.70 $\pm$ 0.24 \\
55006 & 20.71 $\pm$  0.01 & 1.72 $\pm$ 0.37 & 1.60 $\pm$ 0.61 & 3.40 $\pm$ 1.02 & 0.86 $\pm$ 0.29 \\
750374 & 21.21 $\pm$  0.09 & 0.50 $\pm$ 0.09 & 0.33 $\pm$ 0.12 & 2.03 $\pm$ 0.46 & 0.43 $\pm$ 0.17 \\
3173601 & 20.41 $\pm$  0.09 & - & - & - & - \\
365375 & 20.69 $\pm$  0.09 & 1.56 $\pm$ 0.11 & 1.34 $\pm$ 0.01 & 1.05 $\pm$ 0.33 & 0.74 $\pm$ 0.01 \\
\hline
\end{tabular}
\tablefoot{The division line splits the galaxies with stellar mass greater or lower than 8$\times$10$^{10}$M$_{\odot}$. Columns: (1) GAMA ID, (2) Magnitude (3) Effective radius in kpc (4) Circularized effective radius in kpc (5) S\'ersic index (6) Axis ratio.}
\end{table*}

%% file: struct_param_table_i_band.tex
\begin{table*}
\centering
\label{tab:struct_param_i_band}
\caption{List with the $i$-band structural parameters for our MUG sample}
\begin{tabular}{cccccc}
\hline \hline
CATAID & mag $\pm \delta$mag & r$_{\rm e} \pm \delta$r$_{\rm e}$ & r$_{\rm e,circ} \pm \delta$r$_{\rm e,circ}$ & n $\pm~\delta$n & ar $\pm~\delta$ar \\
 & & [kpc] & [kpc] & & \\
\hline
79071 & 16.50 $\pm$  0.01 & 2.14 $\pm$ 0.01 & 1.78 $\pm$ 0.13 & 2.01 $\pm$ 0.05 & 0.69 $\pm$ 0.10 \\
16143 & 18.00 $\pm$  0.01 & 1.67 $\pm$ 0.06 & 1.57 $\pm$ 0.19 & 5.87 $\pm$ 0.19 & 0.88 $\pm$ 0.15 \\
784327 & 19.08 $\pm$  0.12 & 1.18 $\pm$ 0.22 & 1.11 $\pm$ 0.22 & 1.91 $\pm$ 0.44 & 0.89 $\pm$ 0.03 \\
609701 & 19.22 $\pm$  0.09 & 2.01 $\pm$ 0.46 & 1.93 $\pm$ 0.60 & 0.95 $\pm$ 0.75 & 0.92 $\pm$ 0.14 \\
493286 & 19.51 $\pm$  0.09 & 1.68 $\pm$ 0.07 & 1.60 $\pm$ 0.34 & 0.95 $\pm$ 0.62 & 0.91 $\pm$ 0.30 \\
319149 & 18.77 $\pm$  0.05 & 1.78 $\pm$ 0.22 & 1.63 $\pm$ 0.36 & 2.03 $\pm$ 0.46 & 0.83 $\pm$ 0.16 \\
84466 & 18.74 $\pm$  0.19 & 1.99 $\pm$ 0.24 & 1.87 $\pm$ 0.44 & 1.08 $\pm$ 0.82 & 0.89 $\pm$ 0.20 \\
178241 & 18.64 $\pm$  0.40 & 1.91 $\pm$ 0.46 & 1.68 $\pm$ 0.45 & 3.57 $\pm$ 1.13 & 0.77 $\pm$ 0.03 \\
93202 & 17.31 $\pm$  0.01 & 2.45 $\pm$ 0.01 & 2.27 $\pm$ 0.13 & 1.39 $\pm$ 0.09 & 0.86 $\pm$ 0.10 \\
63726 & 19.44 $\pm$  0.05 & 1.56 $\pm$ 0.19 & 1.47 $\pm$ 0.46 & 1.63 $\pm$ 0.45 & 0.88 $\pm$ 0.33 \\
765033 & 18.83 $\pm$  0.01 & 0.68 $\pm$ 0.01 & 0.39 $\pm$ 0.06 & 5.63 $\pm$ 0.13 & 0.33 $\pm$ 0.09 \\
4220443 & 19.44 $\pm$  0.01 & 1.29 $\pm$ 0.27 & 0.95 $\pm$ 0.32 & 3.52 $\pm$ 1.10 & 0.55 $\pm$ 0.13 \\
422365 & 19.10 $\pm$  0.10 & 4.15 $\pm$ 0.53 & 3.71 $\pm$ 0.81 & 2.82 $\pm$ 0.63 & 0.80 $\pm$ 0.15 \\
71471 & 18.49 $\pm$  0.09 & 3.09 $\pm$ 0.08 & 2.92 $\pm$ 0.23 & 1.42 $\pm$ 0.33 & 0.89 $\pm$ 0.09 \\
3873542 & 17.94 $\pm$  0.02 & 1.89 $\pm$ 0.06 & 1.49 $\pm$ 0.33 & 2.86 $\pm$ 0.08 & 0.62 $\pm$ 0.24 \\
300986 & 18.60 $\pm$  0.09 & 0.75 $\pm$ 0.04 & 0.73 $\pm$ 0.09 & 1.93 $\pm$ 0.59 & 0.95 $\pm$ 0.13 \\
221269 & 19.18 $\pm$  0.20 & 2.13 $\pm$ 0.30 & 1.98 $\pm$ 0.58 & 0.70 $\pm$ 0.59 & 0.86 $\pm$ 0.27 \\
288762 & 19.47 $\pm$  0.09 & 1.91 $\pm$ 0.11 & 1.34 $\pm$ 0.26 & 0.77 $\pm$ 0.58 & 0.49 $\pm$ 0.14 \\
138954 & 18.35 $\pm$  0.03 & 1.97 $\pm$ 0.01 & 1.93 $\pm$ 0.29 & 5.12 $\pm$ 0.54 & 0.95 $\pm$ 0.28 \\
855304 & 19.36 $\pm$  0.09 & 1.00 $\pm$ 0.05 & 0.84 $\pm$ 0.08 & 1.27 $\pm$ 0.50 & 0.71 $\pm$ 0.07 \\
791716 & 19.07 $\pm$  0.01 & 0.81 $\pm$ 0.01 & 0.75 $\pm$ 0.11 & 1.97 $\pm$ 0.01 & 0.87 $\pm$ 0.23 \\
693193 & 17.74 $\pm$  0.01 & 1.70 $\pm$ 0.03 & 1.11 $\pm$ 0.46 & 9.67 $\pm$ 0.05 & 0.42 $\pm$ 0.34 \\
\hline \hline
723783 & 17.67 $\pm$  0.08 & 0.91 $\pm$ 0.13 & 0.79 $\pm$ 0.22 & 1.16 $\pm$ 0.11 & 0.76 $\pm$ 0.21 \\
486049 & 19.22 $\pm$  0.20 & 2.85 $\pm$ 0.31 & 2.61 $\pm$ 0.62 & 0.92 $\pm$ 0.71 & 0.84 $\pm$ 0.22 \\
388238 & 18.24 $\pm$  0.09 & - & - & - & - \\
55006 & 19.19 $\pm$  0.01 & 2.74 $\pm$ 0.37 & 2.51 $\pm$ 0.76 & 1.10 $\pm$ 1.02 & 0.84 $\pm$ 0.29 \\
750374 & 18.81 $\pm$  0.09 & 0.79 $\pm$ 0.09 & 0.58 $\pm$ 0.15 & 1.71 $\pm$ 0.46 & 0.53 $\pm$ 0.17 \\
3173601 & 19.11 $\pm$  0.09 & 1.05 $\pm$ 0.06 & 1.03 $\pm$ 0.13 & 1.87 $\pm$ 0.71 & 0.97 $\pm$ 0.12 \\
365375 & 19.94 $\pm$  0.09 & 1.89 $\pm$ 0.11 & 1.83 $\pm$ 0.01 & 0.61 $\pm$ 0.33 & 0.94 $\pm$ 0.01 \\
\hline
\end{tabular}
\tablefoot{The division line splits the galaxies with stellar mass greater or lower than 8$\times$10$^{10}$M$_{\odot}$. Columns: (1) GAMA ID, (2) Magnitude (3) Effective radius in kpc (4) Circularized effective radius in kpc (5) S\'ersic index (6) Axis ratio.}
\end{table*}

%% file: struct_param_table_z_band.tex
\begin{table*}
\centering
\label{tab:struct_param_z_band}
\caption{List with the $Z$-band structural parameters for our MUG sample}
\begin{tabular}{cccccc}
\hline \hline
CATAID & mag $\pm \delta$mag & r$_{\rm e} \pm \delta$r$_{\rm e}$ & r$_{\rm e,circ} \pm \delta$r$_{\rm e,circ}$ & n $\pm~\delta$n & ar $\pm~\delta$ar \\
 & & [kpc] & [kpc] & & \\
\hline
79071 & 16.18 $\pm$  0.03 & 1.69 $\pm$ 0.09 & 1.39 $\pm$ 0.17 & 3.57 $\pm$ 0.24 & 0.68 $\pm$ 0.10 \\
16143 & 17.75 $\pm$  0.02 & 1.02 $\pm$ 0.02 & 0.81 $\pm$ 0.06 & 4.73 $\pm$ 0.17 & 0.63 $\pm$ 0.07 \\
784327 & 18.67 $\pm$  0.12 & 1.33 $\pm$ 0.09 & 1.23 $\pm$ 0.09 & 1.44 $\pm$ 0.15 & 0.85 $\pm$ 0.01 \\
609701 & 18.74 $\pm$  0.01 & 1.65 $\pm$ 0.02 & 1.45 $\pm$ 0.08 & 2.67 $\pm$ 0.02 & 0.77 $\pm$ 0.06 \\
493286 & 19.29 $\pm$  0.28 & 1.33 $\pm$ 0.30 & 1.25 $\pm$ 0.33 & 1.21 $\pm$ 0.51 & 0.88 $\pm$ 0.07 \\
319149 & 18.33 $\pm$  0.04 & 1.83 $\pm$ 0.08 & 1.50 $\pm$ 0.10 & 4.71 $\pm$ 0.26 & 0.68 $\pm$ 0.03 \\
84466 & 18.27 $\pm$  0.06 & 0.91 $\pm$ 0.05 & 0.72 $\pm$ 0.15 & 3.91 $\pm$ 0.78 & 0.63 $\pm$ 0.18 \\
178241 & 18.47 $\pm$  0.01 & 1.93 $\pm$ 0.02 & 1.79 $\pm$ 0.05 & 2.06 $\pm$ 0.01 & 0.86 $\pm$ 0.02 \\
93202 & 17.08 $\pm$  0.01 & 1.90 $\pm$ 0.01 & 1.83 $\pm$ 0.11 & 2.60 $\pm$ 0.23 & 0.93 $\pm$ 0.10 \\
63726 & 18.76 $\pm$  0.03 & 2.01 $\pm$ 0.04 & 1.77 $\pm$ 0.09 & 5.85 $\pm$ 0.34 & 0.78 $\pm$ 0.05 \\
765033 & 18.66 $\pm$  0.01 & 1.24 $\pm$ 0.02 & 0.77 $\pm$ 0.03 & 2.54 $\pm$ 0.02 & 0.38 $\pm$ 0.02 \\
4220443 & 19.06 $\pm$  0.11 & 1.62 $\pm$ 0.23 & 0.91 $\pm$ 0.62 & 3.89 $\pm$ 1.56 & 0.32 $\pm$ 0.34 \\
422365 & 19.13 $\pm$  0.11 & 1.93 $\pm$ 0.28 & 1.82 $\pm$ 0.40 & 3.69 $\pm$ 1.48 & 0.89 $\pm$ 0.14 \\
71471 & 17.95 $\pm$  0.01 & 2.37 $\pm$ 0.01 & 1.84 $\pm$ 0.17 & 5.35 $\pm$ 0.14 & 0.61 $\pm$ 0.11 \\
3873542 & 17.90 $\pm$  0.01 & 1.41 $\pm$ 0.01 & 1.00 $\pm$ 0.19 & 1.88 $\pm$ 0.06 & 0.50 $\pm$ 0.19 \\
300986 & 18.14 $\pm$  0.06 & 1.29 $\pm$ 0.08 & 1.08 $\pm$ 0.31 & 3.32 $\pm$ 0.66 & 0.70 $\pm$ 0.32 \\
221269 & 18.89 $\pm$  0.12 & 2.76 $\pm$ 0.18 & 2.27 $\pm$ 0.18 & 0.44 $\pm$ 0.04 & 0.68 $\pm$ 0.02 \\
288762 & 19.19 $\pm$  0.28 & 1.60 $\pm$ 0.37 & 1.26 $\pm$ 0.31 & 1.15 $\pm$ 0.48 & 0.62 $\pm$ 0.02 \\
138954 & 17.93 $\pm$  0.01 & 0.89 $\pm$ 0.01 & 0.62 $\pm$ 0.10 & 9.59 $\pm$ 0.02 & 0.50 $\pm$ 0.15 \\
855304 & 18.14 $\pm$  0.01 & 2.45 $\pm$ 0.02 & 1.88 $\pm$ 0.40 & 6.52 $\pm$ 0.32 & 0.59 $\pm$ 0.25 \\
791716 & 18.38 $\pm$  0.01 & 2.12 $\pm$ 0.03 & 1.76 $\pm$ 0.06 & 2.94 $\pm$ 0.02 & 0.69 $\pm$ 0.03 \\
693193 & 17.42 $\pm$  0.01 & - & - & - & - \\
\hline \hline
723783 & 17.47 $\pm$  0.01 & 0.67 $\pm$ 0.01 & 0.50 $\pm$ 0.11 & 2.59 $\pm$ 0.23 & 0.56 $\pm$ 0.25 \\
486049 & 18.92 $\pm$  0.12 & 1.64 $\pm$ 0.11 & 1.31 $\pm$ 0.22 & 1.81 $\pm$ 0.18 & 0.64 $\pm$ 0.13 \\
388238 & 17.60 $\pm$  0.01 & - & - & - & - \\
55006 & 18.85 $\pm$  0.01 & 1.43 $\pm$ 0.02 & 1.03 $\pm$ 0.11 & 2.81 $\pm$ 0.02 & 0.52 $\pm$ 0.10 \\
750374 & 18.17 $\pm$  0.01 & - & - & - & - \\
3173601 & 18.97 $\pm$  0.12 & 1.15 $\pm$ 0.08 & 0.69 $\pm$ 0.31 & 0.64 $\pm$ 0.06 & 0.36 $\pm$ 0.27 \\
365375 & 18.97 $\pm$  0.02 & 1.89 $\pm$ 0.02 & 1.22 $\pm$ 0.01 & 2.48 $\pm$ 0.02 & 0.41 $\pm$ 0.01 \\
\hline
\end{tabular}
\tablefoot{The division line splits the galaxies with stellar mass greater or lower than 8$\times$10$^{10}$M$_{\odot}$. Columns: (1) GAMA ID, (2) Magnitude (3) Effective radius in kpc (4) Circularized effective radius in kpc (5) S\'ersic index (6) Axis ratio.}
\end{table*}

%% file: gama_compacts.bbl
\begin{thebibliography}{115}
\expandafter\ifx\csname natexlab\endcsname\relax\def\natexlab#1{#1}\fi

\bibitem[{{Abazajian} {et~al.}(2009){Abazajian}, {Adelman-McCarthy},
  {Ag{\"u}eros}, {Allam}, {Allende Prieto}, {An}, {Anderson}, {Anderson},
  {Annis}, {Bahcall}, \& et~al.}]{Abazajian09}
{Abazajian}, K.~N., {Adelman-McCarthy}, J.~K., {Ag{\"u}eros}, M.~A., {et~al.}
  2009, \apjs, 182, 543

\bibitem[{{Anderson} \& {Darling}(1952)}]{AD52}
{Anderson}, T.~W. \& {Darling}, D.~A. 1952, Ann. Math. Statist., 23, 193

\bibitem[{{Baldry} {et~al.}(2012){Baldry}, {Driver}, {Loveday}, {Taylor},
  {Kelvin}, {Liske}, {Norberg}, {Robotham}, {Brough}, {Hopkins}, {Bamford},
  {Peacock}, {Bland-Hawthorn}, {Conselice}, {Croom}, {Jones}, {Parkinson},
  {Popescu}, {Prescott}, {Sharp}, \& {Tuffs}}]{Baldry12}
{Baldry}, I.~K., {Driver}, S.~P., {Loveday}, J., {et~al.} 2012, \mnras, 421,
  621

\bibitem[{{Baldry} {et~al.}(2018){Baldry}, {Liske}, {Brown}, {Robotham},
  {Driver}, {Dunne}, {Alpaslan}, {Brough}, {Cluver}, {Eardley}, {Farrow},
  {Heymans}, {Hildebrandt}, {Hopkins}, {Kelvin}, {Loveday}, {Moffett},
  {Norberg}, {Owers}, {Taylor}, {Wright}, {Bamford}, {Bland-Hawthorn},
  {Bourne}, {Bremer}, {Colless}, {Conselice}, {Croom}, {Davies}, {Foster},
  {Grootes}, {Holwerda}, {Jones}, {Kafle}, {Kuijken}, {Lara-Lopez},
  {L{\'o}pez-S{\'a}nchez}, {Meyer}, {Phillipps}, {Sutherland}, {van Kampen}, \&
  {Wilkins}}]{Baldry18}
{Baldry}, I.~K., {Liske}, J., {Brown}, M.~J.~I., {et~al.} 2018, \mnras, 474,
  3875

\bibitem[{{Baldry} {et~al.}(2010){Baldry}, {Robotham}, {Hill}, {Driver},
  {Liske}, {Norberg}, {Bamford}, {Hopkins}, {Loveday}, {Peacock}, {Cameron},
  {Croom}, {Cross}, {Doyle}, {Dye}, {Frenk}, {Jones}, {van Kampen}, {Kelvin},
  {Nichol}, {Parkinson}, {Popescu}, {Prescott}, {Sharp}, {Sutherland},
  {Thomas}, \& {Tuffs}}]{Baldry10}
{Baldry}, I.~K., {Robotham}, A.~S.~G., {Hill}, D.~T., {et~al.} 2010, \mnras,
  404, 86

\bibitem[{{Balogh} {et~al.}(1999){Balogh}, {Morris}, {Yee}, {Carlberg}, \&
  {Ellingson}}]{Balogh1999}
{Balogh}, M.~L., {Morris}, S.~L., {Yee}, H.~K.~C., {Carlberg}, R.~G., \&
  {Ellingson}, E. 1999, \apj, 527, 54

\bibitem[{{Barro} {et~al.}(2013){Barro}, {Faber}, {P{\'e}rez-Gonz{\'a}lez},
  {Koo}, {Williams}, {Kocevski}, {Trump}, {Mozena}, {McGrath}, {van der Wel},
  {Wuyts}, {Bell}, {Croton}, {Ceverino}, {Dekel}, {Ashby}, {Cheung},
  {Ferguson}, {Fontana}, {Fang}, {Giavalisco}, {Grogin}, {Guo}, {Hathi},
  {Hopkins}, {Huang}, {Koekemoer}, {Kartaltepe}, {Lee}, {Newman}, {Porter},
  {Primack}, {Ryan}, {Rosario}, {Somerville}, {Salvato}, \& {Hsu}}]{Barro13}
{Barro}, G., {Faber}, S.~M., {P{\'e}rez-Gonz{\'a}lez}, P.~G., {et~al.} 2013,
  \apj, 765, 104

\bibitem[{{Beasley} {et~al.}(2018){Beasley}, {Trujillo}, {Leaman}, \&
  {Montes}}]{Beasley18}
{Beasley}, M.~A., {Trujillo}, I., {Leaman}, R., \& {Montes}, M. 2018, \nat,
  555, 483

\bibitem[{{Bertin} \& {Arnouts}(1996)}]{Bertin1996}
{Bertin}, E. \& {Arnouts}, S. 1996, \aaps, 117, 393

\bibitem[{{Bezanson} {et~al.}(2009){Bezanson}, {van Dokkum}, {Tal},
  {Marchesini}, {Kriek}, {Franx}, \& {Coppi}}]{Bezanson09}
{Bezanson}, R., {van Dokkum}, P.~G., {Tal}, T., {et~al.} 2009, \apj, 697, 1290

\bibitem[{{Bonnarel} {et~al.}(2000){Bonnarel}, {Fernique}, {Bienaym{\'e}},
  {Egret}, {Genova}, {Louys}, {Ochsenbein}, {Wenger}, \&
  {Bartlett}}]{Bonnarel2000}
{Bonnarel}, F., {Fernique}, P., {Bienaym{\'e}}, O., {et~al.} 2000, \aaps, 143,
  33

\bibitem[{{Brough} {et~al.}(2013){Brough}, {Croom}, {Sharp}, {Hopkins},
  {Taylor}, {Baldry}, {Gunawardhana}, {Liske}, {Norberg}, {Robotham}, {Bauer},
  {Bland-Hawthorn}, {Colless}, {Foster}, {Kelvin}, {Lara-Lopez},
  {L{\'o}pez-S{\'a}nchez}, {Loveday}, {Owers}, {Pimbblet}, \&
  {Prescott}}]{Brough13}
{Brough}, S., {Croom}, S., {Sharp}, R., {et~al.} 2013, \mnras, 435, 2903

\bibitem[{{Bruce} {et~al.}(2012){Bruce}, {Dunlop}, {Cirasuolo}, {McLure},
  {Targett}, {Bell}, {Croton}, {Dekel}, {Faber}, {Ferguson}, {Grogin},
  {Kocevski}, {Koekemoer}, {Koo}, {Lai}, {Lotz}, {McGrath}, {Newman}, \& {van
  der Wel}}]{Bruce12}
{Bruce}, V.~A., {Dunlop}, J.~S., {Cirasuolo}, M., {et~al.} 2012, \mnras, 427,
  1666

\bibitem[{{Bruzual} \& {Charlot}(2003)}]{BC03}
{Bruzual}, G. \& {Charlot}, S. 2003, \mnras, 344, 1000

\bibitem[{{Buitrago} {et~al.}(2014){Buitrago}, {Conselice}, {Epinat},
  {Bedregal}, {Gr{\"u}tzbauch}, \& {Weiner}}]{Buitrago14}
{Buitrago}, F., {Conselice}, C.~J., {Epinat}, B., {et~al.} 2014, \mnras, 439,
  1494

\bibitem[{{Buitrago} {et~al.}(2008){Buitrago}, {Trujillo}, {Conselice},
  {Bouwens}, {Dickinson}, \& {Yan}}]{Buitrago08}
{Buitrago}, F., {Trujillo}, I., {Conselice}, C.~J., {et~al.} 2008, \apjl, 687,
  L61

\bibitem[{{Buitrago} {et~al.}(2013){Buitrago}, {Trujillo}, {Conselice}, \&
  {H{\"a}u{\ss}ler}}]{Buitrago13}
{Buitrago}, F., {Trujillo}, I., {Conselice}, C.~J., \& {H{\"a}u{\ss}ler}, B.
  2013, \mnras, 428, 1460

\bibitem[{{Buitrago} {et~al.}(2017){Buitrago}, {Trujillo}, {Curtis-Lake},
  {Montes}, {Cooper}, {Bruce}, {P{\'e}rez-Gonz{\'a}lez}, \&
  {Cirasuolo}}]{Buitrago17}
{Buitrago}, F., {Trujillo}, I., {Curtis-Lake}, E., {et~al.} 2017, \mnras, 466,
  4888

\bibitem[{{Cardelli} {et~al.}(1989){Cardelli}, {Clayton}, \&
  {Mathis}}]{Cardelli1989}
{Cardelli}, J.~A., {Clayton}, G.~C., \& {Mathis}, J.~S. 1989, \apj, 345, 245

\bibitem[{{Carrasco} {et~al.}(2010){Carrasco}, {Conselice}, \&
  {Trujillo}}]{Carrasco10}
{Carrasco}, E.~R., {Conselice}, C.~J., \& {Trujillo}, I. 2010, \mnras, 405,
  2253

\bibitem[{{Cassata} {et~al.}(2010){Cassata}, {Giavalisco}, {Guo}, {Ferguson},
  {Koekemoer}, {Renzini}, {Fontana}, {Salimbeni}, {Dickinson}, {Casertano},
  {Conselice}, {Grogin}, {Lotz}, {Papovich}, {Lucas}, {Straughn}, {Gardner}, \&
  {Moustakas}}]{Cassata10}
{Cassata}, P., {Giavalisco}, M., {Guo}, Y., {et~al.} 2010, \apjl, 714, L79

\bibitem[{{Cassata} {et~al.}(2011){Cassata}, {Giavalisco}, {Guo}, {Renzini},
  {Ferguson}, {Koekemoer}, {Salimbeni}, {Scarlata}, {Grogin}, {Conselice},
  {Dahlen}, {Lotz}, {Dickinson}, \& {Lin}}]{Cassata11}
{Cassata}, P., {Giavalisco}, M., {Guo}, Y., {et~al.} 2011, \apj, 743, 96

\bibitem[{{Chabrier}(2003)}]{Chabrier03}
{Chabrier}, G. 2003, \pasp, 115, 763

\bibitem[{{Charbonnier} {et~al.}(2017){Charbonnier}, {Huertas-Company}, {Gon{\c
  c}alves}, {Men{\'e}ndez-Delmestre}, {Bundy}, {Galliano}, {Moraes}, {Makler},
  {Pereira}, {Erben}, {Hildebrandt}, {Shan}, {Caminha}, {Grossi}, \&
  {Riguccini}}]{Charbonnier17}
{Charbonnier}, A., {Huertas-Company}, M., {Gon{\c c}alves}, T.~S., {et~al.}
  2017, \mnras, 469, 4523

\bibitem[{{Cimatti} {et~al.}(2008){Cimatti}, {Cassata}, {Pozzetti}, {Kurk},
  {Mignoli}, {Renzini}, {Daddi}, {Bolzonella}, {Brusa}, {Rodighiero},
  {Dickinson}, {Franceschini}, {Zamorani}, {Berta}, {Rosati}, \&
  {Halliday}}]{Cimatti08}
{Cimatti}, A., {Cassata}, P., {Pozzetti}, L., {et~al.} 2008, \aap, 482, 21

\bibitem[{{Colless} {et~al.}(2001){Colless}, {Dalton}, {Maddox}, {Sutherland},
  {Norberg}, {Cole}, {Bland-Hawthorn}, {Bridges}, {Cannon}, {Collins}, {Couch},
  {Cross}, {Deeley}, {De Propris}, {Driver}, {Efstathiou}, {Ellis}, {Frenk},
  {Glazebrook}, {Jackson}, {Lahav}, {Lewis}, {Lumsden}, {Madgwick}, {Peacock},
  {Peterson}, {Price}, {Seaborne}, \& {Taylor}}]{2dFGRS}
{Colless}, M., {Dalton}, G., {Maddox}, S., {et~al.} 2001, \mnras, 328, 1039

\bibitem[{{Conselice} {et~al.}(2011){Conselice}, {Bluck}, {Buitrago}, {Bauer},
  {Gr{\"u}tzbauch}, {Bouwens}, {Bevan}, {Mortlock}, {Dickinson}, {Daddi},
  {Yan}, {Scott}, {Chapman}, {Chary}, {Ferguson}, {Giavalisco}, {Grogin},
  {Illingworth}, {Jogee}, {Koekemoer}, {Lucas}, {Mobasher}, {Moustakas},
  {Papovich}, {Ravindranath}, {Siana}, {Teplitz}, {Trujillo}, {Urry}, \&
  {Weinzirl}}]{Conselice11}
{Conselice}, C.~J., {Bluck}, A.~F.~L., {Buitrago}, F., {et~al.} 2011, \mnras,
  413, 80

\bibitem[{{Daddi} {et~al.}(2005){Daddi}, {Renzini}, {Pirzkal}, {Cimatti},
  {Malhotra}, {Stiavelli}, {Xu}, {Pasquali}, {Rhoads}, {Brusa}, {di Serego
  Alighieri}, {Ferguson}, {Koekemoer}, {Moustakas}, {Panagia}, \&
  {Windhorst}}]{Daddi05}
{Daddi}, E., {Renzini}, A., {Pirzkal}, N., {et~al.} 2005, \apj, 626, 680

\bibitem[{{Damjanov} {et~al.}(2014){Damjanov}, {Hwang}, {Geller}, \&
  {Chilingarian}}]{Damjanov14}
{Damjanov}, I., {Hwang}, H.~S., {Geller}, M.~J., \& {Chilingarian}, I. 2014,
  \apj, 793, 39

\bibitem[{{Damjanov} {et~al.}(2009){Damjanov}, {McCarthy}, {Abraham},
  {Glazebrook}, {Yan}, {Mentuch}, {Le Borgne}, {Savaglio}, {Crampton},
  {Murowinski}, {Juneau}, {Carlberg}, {J{\o}rgensen}, {Roth}, {Chen}, \&
  {Marzke}}]{Damjanov09}
{Damjanov}, I., {McCarthy}, P.~J., {Abraham}, R.~G., {et~al.} 2009, \apj, 695,
  101

\bibitem[{{Damjanov} {et~al.}(2015){Damjanov}, {Zahid}, {Geller}, \&
  {Hwang}}]{Damjanov15}
{Damjanov}, I., {Zahid}, H.~J., {Geller}, M.~J., \& {Hwang}, H.~S. 2015, \apj,
  815, 104

\bibitem[{{de Jong} {et~al.}(2017)}]{deJong17}
{de Jong}, J.~T.~A. {et~al.} 2017, \aap, 604, A134

\bibitem[{{de la Rosa} {et~al.}(2016){de la Rosa}, {La Barbera}, {Ferreras},
  {S{\'a}nchez Almeida}, {Dalla Vecchia}, {Mart{\'{\i}}nez-Valpuesta}, \&
  {Stringer}}]{delaRosa16}
{de la Rosa}, I.~G., {La Barbera}, F., {Ferreras}, I., {et~al.} 2016, \mnras,
  457, 1916

\bibitem[{{De Lucia} \& {Blaizot}(2007)}]{deLucia07}
{De Lucia}, G. \& {Blaizot}, J. 2007, \mnras, 375, 2

\bibitem[{{D{\'{\i}}az-Garc{\'{\i}}a}
  {et~al.}(2017){D{\'{\i}}az-Garc{\'{\i}}a}, {Cenarro}, {L{\'o}pez-Sanjuan},
  {Ferreras}, {Cervi{\~n}o}, {Fern{\'a}ndez-Soto}, {M{\'a}rquez}, {Povi{\'c}},
  {San Roman}, {Viironen}, {Moles}, {Crist{\'o}bal-Hornillos}, {Alfaro},
  {Aparicio-Villegas}, {Ben{\'{\i}}tez}, {Broadhurst}, {Cabrera-Ca{\~n}o},
  {Castander}, {Cepa}, {Gonz{\'a}lez Delgado}, {Husillos}, {Infante},
  {Aguerri}, {Masegosa}, {Molino}, {del Olmo}, {Perea}, {Prada}, {Quintana}, \&
  {Mart{\'{\i}}nez}}]{LADG_UVJ}
{D{\'{\i}}az-Garc{\'{\i}}a}, L.~A., {Cenarro}, A.~J., {L{\'o}pez-Sanjuan}, C.,
  {et~al.} 2017, ArXiv e-prints [\eprint[arXiv]{1711.10590}]

\bibitem[{{Driver} {et~al.}(2016){Driver}, {Davies}, {Meyer}, {Power},
  {Robotham}, {Baldry}, {Liske}, \& {Norberg}}]{WAVES}
{Driver}, S.~P., {Davies}, L.~J., {Meyer}, M., {et~al.} 2016, The Universe of
  Digital Sky Surveys, 42, 205

\bibitem[{{Driver} {et~al.}(2011){Driver}, {Hill}, {Kelvin}, {Robotham},
  {Liske}, {Norberg}, {Baldry}, {Bamford}, {Hopkins}, {Loveday}, {Peacock},
  {Andrae}, {Bland-Hawthorn}, {Brough}, {Brown}, {Cameron}, {Ching}, {Colless},
  {Conselice}, {Croom}, {Cross}, {de Propris}, {Dye}, {Drinkwater}, {Ellis},
  {Graham}, {Grootes}, {Gunawardhana}, {Jones}, {van Kampen}, {Maraston},
  {Nichol}, {Parkinson}, {Phillipps}, {Pimbblet}, {Popescu}, {Prescott},
  {Roseboom}, {Sadler}, {Sansom}, {Sharp}, {Smith}, {Taylor}, {Thomas},
  {Tuffs}, {Wijesinghe}, {Dunne}, {Frenk}, {Jarvis}, {Madore}, {Meyer},
  {Seibert}, {Staveley-Smith}, {Sutherland}, \& {Warren}}]{Driver11}
{Driver}, S.~P., {Hill}, D.~T., {Kelvin}, L.~S., {et~al.} 2011, \mnras, 413,
  971

\bibitem[{{Driver} {et~al.}(2013){Driver}, {Robotham}, {Bland-Hawthorn},
  {Brown}, {Hopkins}, {Liske}, {Phillipps}, \& {Wilkins}}]{Driver:13}
{Driver}, S.~P., {Robotham}, A.~S.~G., {Bland-Hawthorn}, J., {et~al.} 2013,
  \mnras, 430, 2622

\bibitem[{{Edge} {et~al.}(2013){Edge}, {Sutherland}, {Kuijken}, {Driver},
  {McMahon}, {Eales}, \& {Emerson}}]{Edge13}
{Edge}, A., {Sutherland}, W., {Kuijken}, K., {et~al.} 2013, The Messenger, 154,
  32

\bibitem[{{Emsellem}(2013)}]{Emsellem13}
{Emsellem}, E. 2013, \mnras, 433, 1862

\bibitem[{{Fabian} {et~al.}(2013){Fabian}, {Sanders}, {Haehnelt}, {Rees}, \&
  {Miller}}]{Fabian13}
{Fabian}, A.~C., {Sanders}, J.~S., {Haehnelt}, M., {Rees}, M.~J., \& {Miller},
  J.~M. 2013, \mnras, 431, L38

\bibitem[{{Fakhouri} \& {Ma}(2010)}]{Fakhouri10}
{Fakhouri}, O. \& {Ma}, C.-P. 2010, \mnras, 401, 2245

\bibitem[{{Ferr{\'e}-Mateu} {et~al.}(2015){Ferr{\'e}-Mateu}, {Mezcua},
  {Trujillo}, {Balcells}, \& {van den Bosch}}]{Ferre-Mateu15}
{Ferr{\'e}-Mateu}, A., {Mezcua}, M., {Trujillo}, I., {Balcells}, M., \& {van
  den Bosch}, R.~C.~E. 2015, \apj, 808, 79

\bibitem[{{Ferr{\'e}-Mateu} {et~al.}(2017){Ferr{\'e}-Mateu}, {Trujillo},
  {Mart{\'{\i}}n-Navarro}, {Vazdekis}, {Mezcua}, {Balcells}, \&
  {Dom{\'{\i}}nguez}}]{Ferre-Mateu17}
{Ferr{\'e}-Mateu}, A., {Trujillo}, I., {Mart{\'{\i}}n-Navarro}, I., {et~al.}
  2017, \mnras, 467, 1929

\bibitem[{{Ferr{\'e}-Mateu} {et~al.}(2013){Ferr{\'e}-Mateu}, {Vazdekis}, \& {de
  la Rosa}}]{Ferre-Mateu13}
{Ferr{\'e}-Mateu}, A., {Vazdekis}, A., \& {de la Rosa}, I.~G. 2013, \mnras,
  431, 440

\bibitem[{{Ferreras} {et~al.}(2012){Ferreras}, {Pasquali}, {Khochfar},
  {Kuntschner}, {K{\"u}mmel}, {Pirzkal}, {Windhorst}, {Malhotra}, {Rhoads},
  {O'Connell}, {Cohen}, {Hathi}, {Ryan}, \& {Yan}}]{Ferreras12}
{Ferreras}, I., {Pasquali}, A., {Khochfar}, S., {et~al.} 2012, \aj, 144, 47

\bibitem[{{Gehrels}(1986)}]{Gehrels1986}
{Gehrels}, N. 1986, \apj, 303, 336

\bibitem[{{Genel} {et~al.}(2018){Genel}, {Nelson}, {Pillepich}, {Springel},
  {Pakmor}, {Weinberger}, {Hernquist}, {Naiman}, {Vogelsberger}, {Marinacci},
  \& {Torrey}}]{Genel18}
{Genel}, S., {Nelson}, D., {Pillepich}, A., {et~al.} 2018, \mnras, 474, 3976

\bibitem[{{Graham} {et~al.}(2015){Graham}, {Dullo}, \& {Savorgnan}}]{Graham15}
{Graham}, A.~W., {Dullo}, B.~T., \& {Savorgnan}, G.~A.~D. 2015, \apj, 804, 32

\bibitem[{{Graham} {et~al.}(2016){Graham}, {Durr{\'e}}, {Savorgnan}, {Medling},
  {Batcheldor}, {Scott}, {Watson}, \& {Marconi}}]{Graham:16}
{Graham}, A.~W., {Durr{\'e}}, M., {Savorgnan}, G.~A.~D., {et~al.} 2016, \apj,
  819, 43

\bibitem[{{Guo} {et~al.}(2013){Guo}, {White}, {Angulo}, {Henriques}, {Lemson},
  {Boylan-Kolchin}, {Thomas}, \& {Short}}]{Guo13}
{Guo}, Q., {White}, S., {Angulo}, R.~E., {et~al.} 2013, \mnras, 428, 1351

\bibitem[{{Guo} {et~al.}(2011){Guo}, {White}, {Boylan-Kolchin}, {De Lucia},
  {Kauffmann}, {Lemson}, {Li}, {Springel}, \& {Weinmann}}]{Guo11}
{Guo}, Q., {White}, S., {Boylan-Kolchin}, M., {et~al.} 2011, \mnras, 413, 101

\bibitem[{{Hopkins} {et~al.}(2009){Hopkins}, {Bundy}, {Murray}, {Quataert},
  {Lauer}, \& {Ma}}]{Hopkins09b}
{Hopkins}, P.~F., {Bundy}, K., {Murray}, N., {et~al.} 2009, \mnras, 398, 898

\bibitem[{Hunter(2007)}]{matplotlib_ref}
Hunter, J.~D. 2007, Computing In Science \& Engineering, 9, 90

\bibitem[{{Johansson} {et~al.}(2012){Johansson}, {Naab}, \&
  {Ostriker}}]{Johansson12}
{Johansson}, P.~H., {Naab}, T., \& {Ostriker}, J.~P. 2012, \apj, 754, 115

\bibitem[{{Kelvin} {et~al.}(2012){Kelvin}, {Driver}, {Robotham}, {Hill},
  {Alpaslan}, {Baldry}, {Bamford}, {Bland-Hawthorn}, {Brough}, {Graham},
  {H{\"a}ussler}, {Hopkins}, {Liske}, {Loveday}, {Norberg}, {Phillipps},
  {Popescu}, {Prescott}, {Taylor}, \& {Tuffs}}]{Kelvin12}
{Kelvin}, L.~S., {Driver}, S.~P., {Robotham}, A.~S.~G., {et~al.} 2012, \mnras,
  421, 1007

\bibitem[{{Kelvin} {et~al.}(2014){Kelvin}, {Driver}, {Robotham}, {Taylor},
  {Graham}, {Alpaslan}, {Baldry}, {Bamford}, {Bauer}, {Bland-Hawthorn},
  {Brown}, {Colless}, {Conselice}, {Holwerda}, {Hopkins}, {Lara-L{\'o}pez},
  {Liske}, {L{\'o}pez-S{\'a}nchez}, {Loveday}, {Norberg}, {Phillipps},
  {Popescu}, {Prescott}, {Sansom}, \& {Tuffs}}]{Kelvin14}
{Kelvin}, L.~S., {Driver}, S.~P., {Robotham}, A.~S.~G., {et~al.} 2014, \mnras,
  444, 1647

\bibitem[{{Khochfar} \& {Silk}(2006)}]{Khochfar06}
{Khochfar}, S. \& {Silk}, J. 2006, \apjl, 648, L21

\bibitem[{{Kuijken} {et~al.}(2015){Kuijken}, {Heymans}, {Hildebrandt},
  {Nakajima}, {Erben}, {de Jong}, {Viola}, {Choi}, {Hoekstra}, {Miller}, {van
  Uitert}, {Amon}, {Blake}, {Brouwer}, {Buddendiek}, {Conti}, {Eriksen},
  {Grado}, {Harnois-D{\'e}raps}, {Helmich}, {Herbonnet}, {Irisarri},
  {Kitching}, {Klaes}, {La Barbera}, {Napolitano}, {Radovich}, {Schneider},
  {Sif{\'o}n}, {Sikkema}, {Simon}, {Tudorica}, {Valentijn}, {Verdoes Kleijn},
  \& {van Waerbeke}}]{Kuijken15}
{Kuijken}, K., {Heymans}, C., {Hildebrandt}, H., {et~al.} 2015, \mnras, 454,
  3500

\bibitem[{{Lange} {et~al.}(2015){Lange}, {Driver}, {Robotham}, {Kelvin},
  {Graham}, {Alpaslan}, {Andrews}, {Baldry}, {Bamford}, {Bland-Hawthorn},
  {Brough}, {Cluver}, {Conselice}, {Davies}, {Haeussler}, {Konstantopoulos},
  {Loveday}, {Moffett}, {Norberg}, {Phillipps}, {Taylor},
  {L{\'o}pez-S{\'a}nchez}, \& {Wilkins}}]{Lange15}
{Lange}, R., {Driver}, S.~P., {Robotham}, A.~S.~G., {et~al.} 2015, \mnras, 447,
  2603

\bibitem[{{Lapi} {et~al.}(2018){Lapi}, {Pantoni}, {Zanisi}, {Shi}, {Mancuso},
  {Massardi}, {Shankar}, {Bressan}, \& {Danese}}]{Lapi18}
{Lapi}, A., {Pantoni}, L., {Zanisi}, L., {et~al.} 2018, \apj, 857, 22

\bibitem[{{Liske} {et~al.}(2015){Liske}, {Baldry}, {Driver}, {Tuffs},
  {Alpaslan}, {Andrae}, {Brough}, {Cluver}, {Grootes}, {Gunawardhana},
  {Kelvin}, {Loveday}, {Robotham}, {Taylor}, {Bamford}, {Bland-Hawthorn},
  {Brown}, {Drinkwater}, {Hopkins}, {Meyer}, {Norberg}, {Peacock}, {Agius},
  {Andrews}, {Bauer}, {Ching}, {Colless}, {Conselice}, {Croom}, {Davies}, {De
  Propris}, {Dunne}, {Eardley}, {Ellis}, {Foster}, {Frenk}, {H{\"a}u{\ss}ler},
  {Holwerda}, {Howlett}, {Ibarra}, {Jarvis}, {Jones}, {Kafle}, {Lacey},
  {Lange}, {Lara-L{\'o}pez}, {L{\'o}pez-S{\'a}nchez}, {Maddox}, {Madore},
  {McNaught-Roberts}, {Moffett}, {Nichol}, {Owers}, {Palamara}, {Penny},
  {Phillipps}, {Pimbblet}, {Popescu}, {Prescott}, {Proctor}, {Sadler},
  {Sansom}, {Seibert}, {Sharp}, {Sutherland}, {V{\'a}zquez-Mata}, {van Kampen},
  {Wilkins}, {Williams}, \& {Wright}}]{Liske15}
{Liske}, J., {Baldry}, I.~K., {Driver}, S.~P., {et~al.} 2015, \mnras, 452, 2087

\bibitem[{{Magorrian} {et~al.}(1998){Magorrian}, {Tremaine}, {Richstone},
  {Bender}, {Bower}, {Dressler}, {Faber}, {Gebhardt}, {Green}, {Grillmair},
  {Kormendy}, \& {Lauer}}]{Magorrian1998}
{Magorrian}, J., {Tremaine}, S., {Richstone}, D., {et~al.} 1998, \aj, 115, 2285

\bibitem[{{Mart{\'{\i}}n-Navarro} {et~al.}(2015){Mart{\'{\i}}n-Navarro},
  {P{\'e}rez-Gonz{\'a}lez}, {Trujillo}, {Esquej}, {Vazdekis}, {Dom{\'{\i}}nguez
  S{\'a}nchez}, {Barro}, {Bruzual}, {Charlot}, {Cava}, {Ferreras}, {Espino},
  {La Barbera}, {Koekemoer}, \& {Cenarro}}]{Martin-Navarro15}
{Mart{\'{\i}}n-Navarro}, I., {P{\'e}rez-Gonz{\'a}lez}, P.~G., {Trujillo}, I.,
  {et~al.} 2015, \apjl, 798, L4

\bibitem[{{McLure} {et~al.}(2013){McLure}, {Pearce}, {Dunlop}, {Cirasuolo},
  {Curtis-Lake}, {Bruce}, {Caputi}, {Almaini}, {Bonfield}, {Bradshaw},
  {Buitrago}, {Chuter}, {Foucaud}, {Hartley}, \& {Jarvis}}]{McLure13a}
{McLure}, R.~J., {Pearce}, H.~J., {Dunlop}, J.~S., {et~al.} 2013, \mnras, 428,
  1088

\bibitem[{{Naab} {et~al.}(2009){Naab}, {Johansson}, \& {Ostriker}}]{Naab09}
{Naab}, T., {Johansson}, P.~H., \& {Ostriker}, J.~P. 2009, \apjl, 699, L178

\bibitem[{{Nipoti} {et~al.}(2012){Nipoti}, {Treu}, {Leauthaud}, {Bundy},
  {Newman}, \& {Auger}}]{Nipoti12}
{Nipoti}, C., {Treu}, T., {Leauthaud}, A., {et~al.} 2012, \mnras, 422, 1714

\bibitem[{{Oke} \& {Gunn}(1983)}]{Oke1983}
{Oke}, J.~B. \& {Gunn}, J.~E. 1983, \apj, 266, 713

\bibitem[{{Oser} {et~al.}(2010{\natexlab{a}}){Oser}, {Ostriker}, {Naab},
  {Johansson}, \& {Burkert}}]{Oser10}
{Oser}, L., {Ostriker}, J.~P., {Naab}, T., {Johansson}, P.~H., \& {Burkert}, A.
  2010{\natexlab{a}}, \apj, 725, 2312

\bibitem[{{Oser} {et~al.}(2010{\natexlab{b}}){Oser}, {Ostriker}, {Naab},
  {Johansson}, \& {Burkert}}]{Oser:10}
{Oser}, L., {Ostriker}, J.~P., {Naab}, T., {Johansson}, P.~H., \& {Burkert}, A.
  2010{\natexlab{b}}, \apj, 725, 2312

\bibitem[{{Peng} {et~al.}(2002){Peng}, {Ho}, {Impey}, \& {Rix}}]{Peng02}
{Peng}, C.~Y., {Ho}, L.~C., {Impey}, C.~D., \& {Rix}, H.-W. 2002, \aj, 124, 266

\bibitem[{{Peng} {et~al.}(2010){Peng}, {Ho}, {Impey}, \& {Rix}}]{Peng10}
{Peng}, C.~Y., {Ho}, L.~C., {Impey}, C.~D., \& {Rix}, H.-W. 2010, \aj, 139,
  2097

\bibitem[{{Peralta de Arriba} {et~al.}(2016){Peralta de Arriba}, {Quilis},
  {Trujillo}, {Cebri{\'a}n}, \& {Balcells}}]{Peralta16}
{Peralta de Arriba}, L., {Quilis}, V., {Trujillo}, I., {Cebri{\'a}n}, M., \&
  {Balcells}, M. 2016, \mnras, 461, 156

\bibitem[{{Poggianti} {et~al.}(2013){Poggianti}, {Calvi}, {Bindoni},
  {D'Onofrio}, {Moretti}, {Valentinuzzi}, {Fasano}, {Fritz}, {De Lucia},
  {Vulcani}, {Bettoni}, {Gullieuszik}, \& {Omizzolo}}]{Poggianti13}
{Poggianti}, B.~M., {Calvi}, R., {Bindoni}, D., {et~al.} 2013, \apj, 762, 77

\bibitem[{{Quilis} \& {Trujillo}(2013)}]{Quilis13}
{Quilis}, V. \& {Trujillo}, I. 2013, \apjl, 773, L8

\bibitem[{{Robitaille} \& {Bressert}(2012)}]{aplpy_ref}
{Robitaille}, T. \& {Bressert}, E. 2012, {APLpy: Astronomical Plotting Library
  in Python}, Astrophysics Source Code Library

\bibitem[{{Robotham} {et~al.}(2010){Robotham}, {Driver}, {Norberg}, {Baldry},
  {Bamford}, {Hopkins}, {Liske}, {Loveday}, {Peacock}, {Cameron}, {Croom},
  {Doyle}, {Frenk}, {Hill}, {Jones}, {van Kampen}, {Kelvin}, {Kuijken},
  {Nichol}, {Parkinson}, {Popescu}, {Prescott}, {Sharp}, {Sutherland},
  {Thomas}, \& {Tuffs}}]{Robotham10}
{Robotham}, A., {Driver}, S.~P., {Norberg}, P., {et~al.} 2010, \pasa, 27, 76

\bibitem[{{Robotham} {et~al.}(2011){Robotham}, {Norberg}, {Driver}, {Baldry},
  {Bamford}, {Hopkins}, {Liske}, {Loveday}, {Merson}, {Peacock}, {Brough},
  {Cameron}, {Conselice}, {Croom}, {Frenk}, {Gunawardhana}, {Hill}, {Jones},
  {Kelvin}, {Kuijken}, {Nichol}, {Parkinson}, {Pimbblet}, {Phillipps},
  {Popescu}, {Prescott}, {Sharp}, {Sutherland}, {Taylor}, {Thomas}, {Tuffs},
  {van Kampen}, \& {Wijesinghe}}]{Robotham11}
{Robotham}, A.~S.~G., {Norberg}, P., {Driver}, S.~P., {et~al.} 2011, \mnras,
  416, 2640

\bibitem[{{Saulder} {et~al.}(2015){Saulder}, {van den Bosch}, \&
  {Mieske}}]{Saulder15}
{Saulder}, C., {van den Bosch}, R.~C.~E., \& {Mieske}, S. 2015, \aap, 578, A134

\bibitem[{{Scharw{\"a}chter} {et~al.}(2016){Scharw{\"a}chter}, {Combes},
  {Salom{\'e}}, {Sun}, \& {Krips}}]{Scharwaechter16}
{Scharw{\"a}chter}, J., {Combes}, F., {Salom{\'e}}, P., {Sun}, M., \& {Krips},
  M. 2016, \mnras, 457, 4272

\bibitem[{{Schlafly} \& {Finkbeiner}(2011)}]{Schlafly:11}
{Schlafly}, E.~F. \& {Finkbeiner}, D.~P. 2011, \apj, 737, 103

\bibitem[{{S{\'e}rsic}(1968)}]{Sersic1968}
{S{\'e}rsic}, J.~L. 1968, {Atlas de galaxias australes}, ed. {Sersic, J.~L.}

\bibitem[{{Shen} {et~al.}(2003){Shen}, {Mo}, {White}, {Blanton}, {Kauffmann},
  {Voges}, {Brinkmann}, \& {Csabai}}]{Shen03}
{Shen}, S., {Mo}, H.~J., {White}, S.~D.~M., {et~al.} 2003, \mnras, 343, 978

\bibitem[{{Shih} \& {Stockton}(2011)}]{Shih11}
{Shih}, H.-Y. \& {Stockton}, A. 2011, \apj, 733, 45

\bibitem[{{Stockton} {et~al.}(2014){Stockton}, {Shih}, {Larson}, \&
  {Mann}}]{Stockton14}
{Stockton}, A., {Shih}, H.-Y., {Larson}, K., \& {Mann}, A.~W. 2014, \apj, 780,
  134

\bibitem[{{Stringer} {et~al.}(2015){Stringer}, {Trujillo}, {Dalla Vecchia}, \&
  {Martinez-Valpuesta}}]{Stringer15}
{Stringer}, M., {Trujillo}, I., {Dalla Vecchia}, C., \& {Martinez-Valpuesta},
  I. 2015, \mnras, 449, 2396

\bibitem[{{Strom} \& {Strom}(1978)}]{Strom:78}
{Strom}, S.~E. \& {Strom}, K.~M. 1978, \apjl, 225, L93

\bibitem[{{Taylor} {et~al.}(2010){Taylor}, {Franx}, {Glazebrook}, {Brinchmann},
  {van der Wel}, \& {van Dokkum}}]{Taylor10}
{Taylor}, E.~N., {Franx}, M., {Glazebrook}, K., {et~al.} 2010, \apj, 720, 723

\bibitem[{{Taylor} {et~al.}(2011){Taylor}, {Hopkins}, {Baldry}, {Brown},
  {Driver}, {Kelvin}, {Hill}, {Robotham}, {Bland-Hawthorn}, {Jones}, {Sharp},
  {Thomas}, {Liske}, {Loveday}, {Norberg}, {Peacock}, {Bamford}, {Brough},
  {Colless}, {Cameron}, {Conselice}, {Croom}, {Frenk}, {Gunawardhana},
  {Kuijken}, {Nichol}, {Parkinson}, {Phillipps}, {Pimbblet}, {Popescu},
  {Prescott}, {Sutherland}, {Tuffs}, {van Kampen}, \& {Wijesinghe}}]{Taylor11}
{Taylor}, E.~N., {Hopkins}, A.~M., {Baldry}, I.~K., {et~al.} 2011, \mnras, 418,
  1587

\bibitem[{{Taylor}(2005)}]{Taylor05}
{Taylor}, M.~B. 2005, in Astronomical Society of the Pacific Conference Series,
  Vol. 347, Astronomical Data Analysis Software and Systems XIV, ed.
  P.~{Shopbell}, M.~{Britton}, \& R.~{Ebert}, 29

\bibitem[{{Thomas} {et~al.}(2003){Thomas}, {Maraston}, \& {Bender}}]{Thomas03}
{Thomas}, D., {Maraston}, C., \& {Bender}, R. 2003, \mnras, 339, 897

\bibitem[{{Tonry} {et~al.}(2000){Tonry}, {Blakeslee}, {Ajhar}, \&
  {Dressler}}]{Tonry00}
{Tonry}, J.~L., {Blakeslee}, J.~P., {Ajhar}, E.~A., \& {Dressler}, A. 2000,
  \apj, 530, 625

\bibitem[{{Tortora} {et~al.}(2016){Tortora}, {La Barbera}, {Napolitano}, {Roy},
  {Radovich}, {Cavuoti}, {Brescia}, {Longo}, {Getman}, {Capaccioli}, {Grado},
  {Kuijken}, {de Jong}, {McFarland}, \& {Puddu}}]{Tortora16}
{Tortora}, C., {La Barbera}, F., {Napolitano}, N.~R., {et~al.} 2016, \mnras,
  457, 2845

\bibitem[{{Tortora} {et~al.}(2018){Tortora}, {Napolitano}, {Spavone}, {La
  Barbera}, {D'Ago}, {Spiniello}, {Kuijken}, {Roy}, {Raj}, {Cavuoti},
  {Brescia}, {Longo}, {Pota}, {Petrillo}, {Radovich}, {Getman}, {Koopmans},
  {Trujillo}, {Verdoes Kleijn}, {Capaccioli}, {Grado}, {Covone},
  {Scognamiglio}, {Blake}, {Glazebrook}, {Joudaki}, {Lidman}, \&
  {Wolf}}]{Tortora18}
{Tortora}, C., {Napolitano}, N.~R., {Spavone}, M., {et~al.} 2018, ArXiv
  e-prints [\eprint[arXiv]{1806.01307}]

\bibitem[{{Trujillo} {et~al.}(2012){Trujillo}, {Carrasco}, \&
  {Ferr{\'e}-Mateu}}]{Trujillo12a}
{Trujillo}, I., {Carrasco}, E.~R., \& {Ferr{\'e}-Mateu}, A. 2012, \apj, 751, 45

\bibitem[{{Trujillo} {et~al.}(2009){Trujillo}, {Cenarro}, {de
  Lorenzo-C{\'a}ceres}, {Vazdekis}, {de la Rosa}, \& {Cava}}]{Trujillo09}
{Trujillo}, I., {Cenarro}, A.~J., {de Lorenzo-C{\'a}ceres}, A., {et~al.} 2009,
  \apjl, 692, L118

\bibitem[{{Trujillo} {et~al.}(2007){Trujillo}, {Conselice}, {Bundy}, {Cooper},
  {Eisenhardt}, \& {Ellis}}]{Trujillo07}
{Trujillo}, I., {Conselice}, C.~J., {Bundy}, K., {et~al.} 2007, \mnras, 382,
  109

\bibitem[{{Trujillo} {et~al.}(2014){Trujillo}, {Ferr{\'e}-Mateu}, {Balcells},
  {Vazdekis}, \& {S{\'a}nchez-Bl{\'a}zquez}}]{Trujillo14}
{Trujillo}, I., {Ferr{\'e}-Mateu}, A., {Balcells}, M., {Vazdekis}, A., \&
  {S{\'a}nchez-Bl{\'a}zquez}, P. 2014, \apjl, 780, L20

\bibitem[{{Trujillo} {et~al.}(2011){Trujillo}, {Ferreras}, \& {de La
  Rosa}}]{I3}
{Trujillo}, I., {Ferreras}, I., \& {de La Rosa}, I.~G. 2011, \mnras, 415, 3903

\bibitem[{{Trujillo} {et~al.}(2006){Trujillo}, {Feulner}, {Goranova}, {Hopp},
  {Longhetti}, {Saracco}, {Bender}, {Braito}, {Della Ceca}, {Drory},
  {Mannucci}, \& {Severgnini}}]{Trujillo06a}
{Trujillo}, I., {Feulner}, G., {Goranova}, Y., {et~al.} 2006, \mnras, 373, L36

\bibitem[{{Valentinuzzi} {et~al.}(2010{\natexlab{a}}){Valentinuzzi}, {Fritz},
  {Poggianti}, {Cava}, {Bettoni}, {Fasano}, {D'Onofrio}, {Couch}, {Dressler},
  {Moles}, {Moretti}, {Omizzolo}, {Kj{\ae}rgaard}, {Vanzella}, \&
  {Varela}}]{Valentinuzzi10a}
{Valentinuzzi}, T., {Fritz}, J., {Poggianti}, B.~M., {et~al.}
  2010{\natexlab{a}}, \apj, 712, 226

\bibitem[{{Valentinuzzi} {et~al.}(2010{\natexlab{b}}){Valentinuzzi},
  {Poggianti}, {Saglia}, {Arag{\'o}n-Salamanca}, {Simard},
  {S{\'a}nchez-Bl{\'a}zquez}, {D'onofrio}, {Cava}, {Couch}, {Fritz}, {Moretti},
  \& {Vulcani}}]{Valentinuzzi10b}
{Valentinuzzi}, T., {Poggianti}, B.~M., {Saglia}, R.~P., {et~al.}
  2010{\natexlab{b}}, \apjl, 721, L19

\bibitem[{{van den Bosch} {et~al.}(2012){van den Bosch}, {Gebhardt},
  {G{\"u}ltekin}, {van de Ven}, {van der Wel}, \& {Walsh}}]{vandenBosch12}
{van den Bosch}, R.~C.~E., {Gebhardt}, K., {G{\"u}ltekin}, K., {et~al.} 2012,
  \nat, 491, 729

\bibitem[{{van der Wel} {et~al.}(2014){van der Wel}, {Franx}, {van Dokkum},
  {Skelton}, {Momcheva}, {Whitaker}, {Brammer}, {Bell}, {Rix}, {Wuyts},
  {Ferguson}, {Holden}, {Barro}, {Koekemoer}, {Chang}, {McGrath},
  {H{\"a}ussler}, {Dekel}, {Behroozi}, {Fumagalli}, {Leja}, {Lundgren},
  {Maseda}, {Nelson}, {Wake}, {Patel}, {Labb{\'e}}, {Faber}, {Grogin}, \&
  {Kocevski}}]{VanderWel14}
{van der Wel}, A., {Franx}, M., {van Dokkum}, P.~G., {et~al.} 2014, \apj, 788,
  28

\bibitem[{{van der Wel} {et~al.}(2011){van der Wel}, {Rix}, {Wuyts}, {McGrath},
  {Koekemoer}, {Bell}, {Holden}, {Robaina}, \& {McIntosh}}]{vanderWel11}
{van der Wel}, A., {Rix}, H.-W., {Wuyts}, S., {et~al.} 2011, \apj, 730, 38

\bibitem[{{van Dokkum} {et~al.}(2010){van Dokkum}, {Whitaker}, {Brammer},
  {Franx}, {Kriek}, {Labb{\'e}}, {Marchesini}, {Quadri}, {Bezanson},
  {Illingworth}, {Muzzin}, {Rudnick}, {Tal}, \& {Wake}}]{vanDokkum10}
{van Dokkum}, P.~G., {Whitaker}, K.~E., {Brammer}, G., {et~al.} 2010, \apj,
  709, 1018

\bibitem[{{Vazdekis} {et~al.}(2012){Vazdekis}, {Ricciardelli}, {Cenarro},
  {Rivero-Gonz{\'a}lez}, {D{\'{\i}}az-Garc{\'{\i}}a}, \&
  {Falc{\'o}n-Barroso}}]{Vazdekis12}
{Vazdekis}, A., {Ricciardelli}, E., {Cenarro}, A.~J., {et~al.} 2012, \mnras,
  424, 157

\bibitem[{{Viola} {et~al.}(2015){Viola}, {Cacciato}, {Brouwer}, {Kuijken},
  {Hoekstra}, {Norberg}, {Robotham}, {van Uitert}, {Alpaslan}, {Baldry},
  {Choi}, {de Jong}, {Driver}, {Erben}, {Grado}, {Graham}, {Heymans},
  {Hildebrandt}, {Hopkins}, {Irisarri}, {Joachimi}, {Loveday}, {Miller},
  {Nakajima}, {Schneider}, {Sif{\'o}n}, \& {Verdoes Kleijn}}]{Viola15}
{Viola}, M., {Cacciato}, M., {Brouwer}, M., {et~al.} 2015, \mnras, 452, 3529

\bibitem[{{Wellons} {et~al.}(2016){Wellons}, {Torrey}, {Ma}, {Rodriguez-Gomez},
  {Pillepich}, {Nelson}, {Genel}, {Vogelsberger}, \& {Hernquist}}]{Wellons16}
{Wellons}, S., {Torrey}, P., {Ma}, C.-P., {et~al.} 2016, \mnras, 456, 1030

\bibitem[{{Williams} {et~al.}(2009){Williams}, {Quadri}, {Franx}, {van Dokkum},
  \& {Labb{\'e}}}]{Williams09}
{Williams}, R.~J., {Quadri}, R.~F., {Franx}, M., {van Dokkum}, P., \&
  {Labb{\'e}}, I. 2009, \apj, 691, 1879

\bibitem[{{Worthey} \& {Ottaviani}(1997)}]{Worthey1997}
{Worthey}, G. \& {Ottaviani}, D.~L. 1997, \apjs, 111, 377

\bibitem[{{Y{\i}ld{\i}r{\i}m} {et~al.}(2015){Y{\i}ld{\i}r{\i}m}, {van den
  Bosch}, {van de Ven}, {Husemann}, {Lyubenova}, {Walsh}, {Gebhardt}, \&
  {G{\"u}ltekin}}]{Yildirim15}
{Y{\i}ld{\i}r{\i}m}, A., {van den Bosch}, R.~C.~E., {van de Ven}, G., {et~al.}
  2015, \mnras, 452, 1792

\bibitem[{{Y{\i}ld{\i}r{\i}m} {et~al.}(2017){Y{\i}ld{\i}r{\i}m}, {van den
  Bosch}, {van de Ven}, {Mart{\'{\i}}n-Navarro}, {Walsh}, {Husemann},
  {G{\"u}ltekin}, \& {Gebhardt}}]{Yildirim17}
{Y{\i}ld{\i}r{\i}m}, A., {van den Bosch}, R.~C.~E., {van de Ven}, G., {et~al.}
  2017, \mnras, 468, 4216

\bibitem[{{Zolotov} {et~al.}(2015){Zolotov}, {Dekel}, {Mandelker}, {Tweed},
  {Inoue}, {DeGraf}, {Ceverino}, {Primack}, {Barro}, \& {Faber}}]{Zolotov15}
{Zolotov}, A., {Dekel}, A., {Mandelker}, N., {et~al.} 2015, \mnras, 450, 2327

\bibitem[{{Zwicky} \& {Kowal}(1968)}]{ZwKow:68}
{Zwicky}, F. \& {Kowal}, C.~T. 1968, {``Catalogue of Galaxies and of Clusters
  of Galaxies'', Volume VI}

\end{thebibliography}
